%% file: main.tex
\input{header}

\title{Optimal Transport of A Free Quantum Particle and its Shape Space Interpretation}

\author{
	Bernadette Lessel \\
	Universität Bonn\\
	\texttt{blessel@uni-bonn.de} \\
}

\begin{document}
	\maketitle
	
	\begin{abstract}
		A solution of the free Schrödinger equation is investigated by means of Optimal transport. The curve of probability measures $\mu_t$ this solution defines is shown to be an absolutely continuous curve in the Wasserstein space $W_2(\R^3)$. The optimal transport map from $\mu_t$ to $\mu_s$, the cost for this transport (i.e. the Wasserstein distance) and the value of the Fisher information along $\mu_t$ are being calculated. It is finally shown that this solution of the free Schrödinger equation can naturally be interpreted as a curve in so-called Shape space, which forgets any positioning in space but only describes properties of shapes. In Shape space, $\mu_t$ continues to be a shortest path geodesic.
	\end{abstract}

\tableofcontents

\section{Introduction}
%---------------------
The probability interpretation of quantum mechanics allows for an investigation of solutions $\Psi$ of the, in our case, free Schrödinger equation by means of optimal transport. After $\Psi(x,t)$ is identified with the probability measure $\mu_t:=|\psi(x,t)|^2dx^3$ one can begin to wonder how $\mu_t$ may be transported optimally onto $\mu_s$ and what the cost for this optimal transport may be. This article shows that for a specific solution of the free Schrödinger equation, the optimal transport map is nothing but the (time-dependent) flow $F$ of the vector field $\frac{1}{m}$$\nabla S$, where $S$ is the phase of the wave function. In particular, $\mu_t=F_s(\cdot,t)_\#\mu_s$. The paper further shows that $\mu_t$ is an absolutely continuous curve in Wasserstein space, with $\frac{1}{m}$$\nabla S$ being its accompanying vector field. Along this curve, the Fisher Information is shown to be monotonously decreasing. 

In the second part of the paper it is analysed which properties $\mu_t$ has when being projected onto Shape space. Defined in \cite{Lessel2018,Lessel2025a}, the so-called Shape space considers only the property of \emph{shape} of a measure, while forgetting entirely about its location in space. After repeating in subsection \ref{s.shapespaces} the definition and some of the most relevant properties of Shape space, it is shown that $\mu_t$ retains its most important properties in this respect after its projection onto Shape space. In particular, it remains a geodesic in the sense of shortest paths. Also, the notion of a tangent space on Shape space proposed in \cite{Lessel2018,Lessel2025a} is used here to show that with respect to this definition, $\frac{1}{m}$$\nabla S$ can naturally be regarded as staying tangent to $\mu_t$ also after its projection onto Shape space. It is thus suggested that Shape spaces may be an appropriate structure to investige essential properties of free quantum evolution.

This paper is the third and final spin-off from the PhD dissertation \cite{Lessel2018}, following \cite{Lessel2020} and \cite{Lessel2025a}. They all share an introduction to Optimal transport and Wasserstein geometry that essentially only differs by the extent of detail relevant for the respective topic of the article.

\input{Kapitel/WassersteinSpaces}
\input{Kapitel/QuantumDynamics}
\input{Kapitel/DynamicsShapeSpace}

\section*{Acknowledgements}
This article is a spin-off from my PhD dissertation \cite{Lessel2018}, which I wrote being part of the DFG Research Training Group 1493 “Mathematical Structures in Modern Quantum Physics“, Georg-August-Universität Göttingen. I am indebted to my PhD supervisor Prof. Thomas Schick for his guidance and mathematical advice throughout my PhD.

\bibliographystyle{alpha}       % Set the bibliography style to AMS plain alphabetized. (Can use ``amsalpha'', ``abbrv'', "apalike", acm)
\bibliography{da,C:/Users/Admin/Documents/Archives/Bibliothek/Mathe.bib,C:/Users/Admin/Documents/Archives/Bibliothek/Physics.bib}

\end{document}

%% file: header.tex
\documentclass{article}

\usepackage{arxiv}

\usepackage[utf8]{inputenc} % allow utf-8 input
\usepackage[T1]{fontenc}    % use 8-bit T1 fonts
\usepackage{hyperref}       % hyperlinks
\usepackage{url}            % simple URL typesetting
\usepackage{booktabs}       % professional-quality tables
\usepackage{amsfonts}       % blackboard math symbols
\usepackage{nicefrac}       % compact symbols for 1/2, etc.
\usepackage{microtype}      % microtypography
\usepackage{lipsum}		% Can be removed after putting your text content

\usepackage{tikz}                        % For commutative diagrams
\usetikzlibrary{matrix,arrows}
\usepackage{amsmath}
\usepackage{amsthm}
\usepackage{amssymb}
\usepackage{mathtools}                  % For the big | 
\usepackage{dsfont}                     % For the identity matrix
\usepackage{amsxtra}                     % Use various AMS packages
\usepackage{epsfig}
\usepackage{verbatim}
\usepackage{epigraph}                    % For epigraphs...
\DeclarePairedDelimiter\abs{\lvert}{\rvert}%
\theoremstyle{plain}
\newtheorem{thm}{Theorem} %Hier und im Folgendem theorem -->thm
\newtheorem{prop}[thm]{Proposition}
\newtheorem{corollary}[thm]{Corollary}
\newtheorem{lemma}[thm]{Lemma}

\theoremstyle{definition}
\newtheorem{dfn}[thm]{Definition}

\newtheorem{example}[thm]{Example}

\theoremstyle{remark}
\newtheorem{remark}[thm]{Remark}
\newcommand{\bigslant}[2]{{\raisebox{.2em}{$#1$}\left/\raisebox{-.2em}{$#2$}\right.}}
\newcommand{\be}{\begin{equation}}
\newcommand{\ee}{\end{equation}}
\newcommand{\bdm}{\begin{displaymath}}
\newcommand{\edm}{\end{displaymath}}
\newcommand{\ba}[1]{\begin{array}{#1}}
\newcommand{\ea}{\end{array}}

\newcommand{\bcen}{\begin{center}}
\newcommand{\ecen}{\end{center}}
\newcommand{\btab}{\begin{tabular}}
\newcommand{\etab}{\end{tabular}}

\newcommand{\ra}{\rightarrow}
\newcommand{\lra}{\longrightarrow}

\newcommand{\Lra}{\Leftrightarrow}
\newcommand{\lmapsto}{\longmapsto}

\newcommand{\<}{\langle}
\renewcommand{\>}{\rangle} 
\newcommand{\vertiii}[1]{{\left\vert\kern-0.25ex\left\vert\kern-0.25ex\left\vert #1 
		\right\vert\kern-0.25ex\right\vert\kern-0.25ex\right\vert}}
\newcommand{\PM}{\ensuremath{\mathcal{P}}} % Menge der W'Maße
\newcommand{\B}{\ensuremath{\mathcal{B}}} % Borel-Sigma-Algebra
\newcommand{\Sh}{\ensuremath{\mathcal{S}}} % Space of Shapes

\newcommand{\R}{\ensuremath{\mathbb{R}}}
\newcommand{\N}{\ensuremath{\mathbb{N}}}
\newcommand{\C}{\ensuremath{\mathbb{C}}}

%% file: Kapitel/WassersteinSpaces.tex
%-------------------------------------------
\section{Wasserstein geometry}\label{ch.was}
%-------------------------------------------
%
We begin this article with an introductory section on Wasserstein geometry, which is a condensed version of \cite{Lessel2018}. In particular, the material included here is standard within the theory of Wasserstein geometry.
The main references in this section are \cite{User}, \cite{GFlows}, \cite{Gigithesis}, \cite{Gigli}, \cite{Vilsmall} and \cite{Vilbig}.

Throughout this article let $\PM(X)$ be the set of probability measures on the topological space $X$, with respect to the Borel $\sigma$-algebra $\B(X)$. A measurable map between two measurable spaces $T:(X,\B(X))\rightarrow (Y,\B(Y))$ induces a map between the respective spaces of probability measures via the \emph{pushforward} $T_\#$ of measures: $T_\#:\PM(X)\rightarrow\PM(Y)$, $\mu\mapsto T_\#\mu$, where $T_\#\mu(A):=\mu(T^{-1}(A))$, for $A\in\B(Y)$. The \emph{support} of a measure $\mu$ is defined by $supp(\mu):=\{x\in X\mid \text{ every open neighbourhood of $x$ has positive $\mu$-measure}\}$. The Lebesgue measure on $\R^n$ is denoted by $\lambda$. 
%
%---------------------------------------------------
\subsection{Optimal transport}\label{sec.was}
%---------------------------------------------------
%
Let $\mu\in\PM(X)$ and $\nu\in\PM(Y)$ be probability measures. A natural question is how to \emph{couple} $\mu$ and $\nu$, i.e. how to relate them with each other. One possibility is to couple them with the help of a measurable map $T:X\rightarrow Y$, namely such that $T_\#\mu=\nu$. However, such a $T$ cannot always be found. This is the case, for example, whenever $\mu$ is a Dirac measure and $\nu$ is not (maps cannot "split mass"). 
A further idea is to try to see $\mu$ and $\nu$ as two sides of the same thing, so to say. This is by looking at the elements of the set 
\begin{equation}\nonumber
Adm(\mu,\nu):=\lbrace \gamma\in\PM(X\times Y)\mid\pi^X_\#\gamma=\mu, \pi^Y_\#\gamma=\nu\rbrace,
\end{equation}
the \emph{admissible plans} between $\mu$ and $\nu$. Here, $\pi^X:X\times Y\rightarrow X$ is the projection onto the $X$-component, i.e. $\pi^X(x,y)=x$. Similarly $\pi^Y$. $Adm(\mu,\nu)$ is never empty, since the product measure $\mu\otimes\nu$ is always an element. And in case there is a map $T$ like above, $\gamma=(Id,T)_\#\mu\in Adm(\mu,\nu)$. So any coupling in terms of maps can be seen as a coupling in terms of admissible plans.

Since $Adm(\mu,\nu)$ is not just not empty but in general has more than one element (for example if $\mu$ and $\nu$ are the sum of $n$ Dirac measures), the question regarding the \emph{best} coupling arises. Of course, a priori it is not clear what "best" actually means. Our perspective is that a coupling should be interpreted as a plan telling how to, instantaneously, rearrange $\mu$ such that it yields $\nu$. Or, put differently, as a plan encoding how to \emph{transport}, $\mu$ onto $\nu$. In this interpretation we can think of $\gamma(A\times B)$ as being the amount of mass which is transported from $A$ to $B$, where according to the definition of $\gamma$, $\gamma(A\times Y)=\mu(A)$ and $\gamma(X \times B)=\nu(B)$ for $A\in\B(X),B\in\B(Y)$.

To make precise what a best element should provide, we assume that we have further data which already relates $X$ and $Y$ with each other. Namely, we assume we have given a measurable function $c:X\times Y\rightarrow\R$. In our interpretation, the number $c(x,y)$ says how much it \emph{costs} to transport one unit of something from $x\in X$ to $y\in Y$. Accordingly, we call $c$ the \emph{cost function}. 
The least cost for transporting $\mu$ to $\nu$ is then given by 

\begin{equation}\label{eq.kant}
C(\mu,\nu):=\inf_{\gamma\in Adm(\mu,\nu)}\int_{X\times Y} c(x,y)\ d\gamma(x,y).
\end{equation}

Thus, a transport plan $\gamma_{opt}\in Adm(\mu,\nu)$ can be considered to be the best plan, or to be \emph{optimal}, in case $C(\mu,\nu)= \int_{X\times Y} c(x,y)\ d\gamma_{opt}(x,y)$. The plan $\mu\otimes\nu$ can be seen as the most inefficient plan, since mass is brought from each measurable subset of positive measure of $X$ to each measurable subset of positive measure of $Y$: $\mu\otimes\nu(A\times B)=\mu(A)\cdot\nu(B)$.
In case $\gamma\in Adm(\mu,\nu)$ is induced by a measurable map $T:X\rightarrow Y$, i.e. in case $\gamma=(Id,T)_\#\mu$, $T$ is called \emph{transport map} and the respective transportation cost is given by $\int_{X\times Y} c(x,T(x))\ d\mu(x)$. The optimization problem

\begin{equation}\nonumber
\widetilde{C}(\mu,\nu):=\inf_{T}\int_{X} c(x,T(x))\ d\mu(x),
\end{equation}

where $T:X\rightarrow Y$ is a measurable map such that $T_\#\mu=\nu$ is called the \emph{Monge formulation} of Optimal transport (\cite{Monge}), whereas \eqref{eq.kant} is called the \emph{Kantorovich formulation} (\cite{Kantorovich}).

Minimizer for \eqref{eq.kant} already exist under mild assumptions on $c$, as we will see in Theorem \ref{thm.min}. For this, we need to introduce Polish spaces.

\begin{dfn}[\emph{\textbf{Metric distance}}]
	A \emph{metric distance}, or just \emph{metric}, on a space $X$ is a map $d:X\times X\rightarrow \R_{\geq0}$ which satisfies the three  conditions $d(x,y)=0$ if and only if $x=y$, $d(x,y)=d(y,x)$ and $d(x,y)\leq d(x,z)+d(z,y)$ for all $x,y,z\in X$.
\end{dfn}

The open balls $B(x,r):=\{y\in X \mid d(x,y)<r\}$ form a base for a topology on $X$, turning $X$ into a topological space. We call this topology the \emph{topology induced by $d$}.

\begin{dfn}[\emph{\textbf{Completely metrizable space}}]
	A topological space $X$ is called \emph{completely metrizable} if there exists at least one metric $d$ on $X$ which induces the given topology on $X$ and which is such that $(X,d)$ is a complete metric space. 
\end{dfn}

\begin{dfn}[\emph{\textbf{Polish space}}]\label{df.polish}
	 A \emph{Polish space} is a separable topological space $X$ which is completely metrizable.
\end{dfn}

When we say that $(X,d)$ is a Polish space, we mean that $X$ is a Polish space and $d$ is a metric on $X$ that induces a topology which coincides with the topology of $X$ and is such that $(X,d)$ is a complete metric space. Such a metric $d$ is called to \emph{metrize} the Polish topology. 

\begin{thm}[\emph{\textbf{Existence of a minimizer}}]\label{thm.min}
Let $X$ and $Y$ be Polish spaces and $c:X\times Y\rightarrow \R$ be a lower semicontinuous cost function such that $c(x,y)\leq a(x)+b(y)\ \forall (x,y)\in X\times Y$ for upper semicontinuous functions $a:X\ra \R\cup\{-\infty\},\ b:Y\ra \R\cup\{-\infty\}$ such that $a\in L^1(\mu),\ b\in L^1(\nu)$.
Then there is an element in $Adm(\mu,\nu)$ which minimizes the Kantorovich formulation of Optimal transport.
\end{thm}

See for example $\cite{Vilbig}$ for a proof. The idea there is to show that $\int c\ d\gamma$ is a lower semicontinuous function on a compact set.

For us, the most important cost functions will be the metrics $d$ which metrize the Polish space under consideration. In this case, of course, $X=Y$. With respect to their induced topology, metrics are continuous and they are bounded from below with $a=b=0$. 

There is a very important theorem expressing when a plan $\gamma\in\PM(X\times Y)$ is optimal for its marginals. To be able to formulate it, we need to introduce some further notions. Again we will not detail the argumentation.

\begin{dfn}[\emph{\textbf{$c$-cyclical monotone set}}]
	A set $Z\subset X\times Y$ is called \emph{$c$-cyclically monotone} if for each $N\in\N$ and each subset $\{(x_i,y_i)\}_{1\leq i\leq N}\subset Z$ of $Z$ containing $N$ elements, it is 
		\begin{equation}\nonumber
		\sum_{i=1}^{N}c(x_i,y_i)\leq \sum_{i=1}^{N}c(x_i,y_{\sigma(i)}), 
		\end{equation}
	for every permutation of the set $\{1,...,N\}$.
\end{dfn}

\begin{dfn}[\emph{\textbf{$c_+$-concavity}}]
		The \emph{$c_+$-transform} of a function $\psi:Y\rightarrow\R\cup\{\pm\infty\}$ is the function $\psi^{c_+}$ defined by
		\begin{eqnarray}\nonumber 	
		\psi^{c_+}: X &\longrightarrow& \R\cup\{-\infty\}\\\nonumber
		\ \ \ \ \ \ \ x  &\longmapsto & \inf_{y\in Y} c(x,y)-\psi(y).
 		\end{eqnarray}
 	A function $\varphi:X\rightarrow\R\cup\{-\infty\}$ is called \emph{$c$-concave} if it is the $c_+$-transform of another function $\psi:Y\rightarrow \R\cup\{-\infty\}$, i.e. if $\varphi=\psi^{c_+}$.
\end{dfn}

\begin{dfn}[\emph{\textbf{$c$-superdifferential}}]
	For a $c$-concave function $\varphi:X\rightarrow\R\cup\{-\infty\}$, the $c$-superdifferential $\partial^{c_+}\varphi\subset X\times Y$ is defined by 
	\begin{equation}\nonumber
	\partial^{c_+}\varphi:=\{(x,y)\in X\times Y \mid \varphi(x)+\varphi^{c_+}(y)=c(x,y)\}.
	\end{equation}
	The $c$-superdifferential at $x\in X$ is the set $\partial^{c_+}\varphi(x):=\{y\in Y\mid (x,y)\in \partial^{c_+}\varphi\}$.
\end{dfn}
The following characterization will be important for us in Section \ref{sec.otquant}.
\begin{prop}\label{prop.later}
	Let $X=Y=\R^n$ and $c(x,y)=\|x-y\|^2/2$. A function $\varphi:\R^d\rightarrow\R\cup\{-\infty\}$ is $c$-concave if and only if the map $\bar{\varphi}(x):=\|x\|^2/2-\varphi(x)$ is convex and lower semicontinuous. In this case, $y\in \partial^{c_+}\varphi(x)$ if and only if $y\in\partial^-\bar{\varphi}(x)$, where $\partial^-$ denotes the usual subdifferential from convex calculus.
\end{prop}

Now we cite from \cite{User} the so called \emph{Fundamental theorem of Optimal transport}.

\begin{thm}[\emph{\textbf{Fundamental theorem of Optimal transport}}]\label{thm.fundot}
	Let the cost function $c:X\times Y\rightarrow\R$ be continuous and bounded from below. Assume further that the measures $\mu\in\PM(X)$ and $\nu\in\PM(Y)$ are such that 
	\begin{equation}\label{eq.condition}
	c(x,y)\leq a(x)+b(y),
	\end{equation}
    for some functions $a\in L^1(\mu)$ and $b\in L^1(\nu).$ For $\gamma\in Adm(\mu,\nu)$ the following three statements are equivalent:
	\begin{itemize}
		\item[1) ]  The plan $\gamma$ is optimal.
		\item[2) ] The set $supp(\gamma)$ is $c$-cyclically monotone.
		\item[3) ] There exists a $c$-concave function $\varphi$ such that $\max\{\varphi,0\}\in L^1(\mu)$ and $supp(\gamma)\subset\partial^{c_+}\varphi$.
	\end{itemize}
\end{thm}
Consequently, optimality depends only on the support of the plan $\gamma$, not on the distribution of mass. If $\gamma$ is optimal for its marginals and $\tilde{\gamma}\in\PM(X\times Y)$ is such that $supp(\tilde{\gamma})\subset supp(\gamma)$, then $\tilde{\gamma}$ is optimal, too, for its marginals.

\begin{remark}\label{rem.later}
Let $T:X\rightarrow Y$ be a map with $T(x)\in\partial^{c_+}\varphi(x)$ for a $c$-concave function $\varphi$, for all $x\in X$. Then for every $\mu\in\PM_2(X)$ such that condition (\ref{eq.condition}) is satisfied for $\nu=T_\#\mu$, the map $T$ is optimal between $\mu$ and $T_\#\mu$. 
\end{remark}

\begin{remark}
	The notions $c$-cyclical monotonicity, $c_+$-concavity and $c$-superdif-ferential generalize notions known from convex analysis: For $X=Y=\R^n$ and $c(x,y)=\<x,y\>$ the Euclidean scalar product, a set is $c$-cyclical monotone if and only if it is cyclically monotone. A function is $c$-convex if and only if it is convex and lower semicontinuous and the $c$-subdifferential is the known subdifferential.
\end{remark}

Next, one can be curious about when an optimal plan $\gamma$ is actually induced by a map, i.e. when $\gamma=(Id,T)_\#\mu$ with $\mu$ being such that $\pi^1_\#\gamma=\mu$. One can show (\cite{User}) that $\gamma$ is induced by a map if and only if there exists a $\gamma$-measurable set $\Gamma\subset X\times Y$ on which $\gamma$ is concentrated, such that for $\mu$-a.e. $x$ there exists only one $y=T(x)\in Y$ such that $(x,y)\in\Gamma$. In this case, $\gamma$ is induced by the map $T$. Since we know from Theorem \ref{thm.fundot} that for optimal $\gamma$ $supp(\gamma)$ is a subset of the $c$-superdifferential of a $c$-concave function $\varphi$, it is necessary to understand in which cases the $c$-superdifferential is single valued.
As in \cite{User}, we will give an answer to this for the cases $X=Y=\R^n$, $c(x,y)=\|x-y\|^2/2$ and $X=Y=M$, $c=d^2/2$, where $M$ is a connected, complete smooth Riemannian manifold and $d$ the corresponding Riemannian metric distance. In both cases, the characterization of the situation in which $\gamma$ is induced by a map holds for so called \emph{regular measures}, which we want to introduce first.

\begin{dfn}[\emph{\textbf{c-c hypersurface}}]
	A subset $A\subset\R^n$ is called a \emph{convex-convex hypersurface} (c-c hypersurface), whenever there exists convex functions $f,g:\R^{n-1}\rightarrow\R$ such that
\begin{equation}\nonumber
A=\{(y,t)\in\R^{n-1}\times\R \mid t=f(y)-g(y)\}.
\end{equation}	
\end{dfn}

\begin{dfn}{\emph{\textbf{(Regular measure)}}}
	A measure $\mu\in\PM(\R^n)$ is called \emph{regular}, in case $\mu(A)=0$ for every c-c hypersurface $A\subset\R^n$.
\end{dfn}

Measures which are absolutely continuous with respect to the Lebesgue measure are, for example, regular.

The following theorem is due to Yann Brenier (\cite{Brenier1987}, \cite{Brenier1991}).
\begin{thm}[\emph{\textbf{Existence of optimal maps}}]\label{thm.Rmap}
	In case $\mu\in\PM(\R^n)$ is such that $\int|x|^2\ d\mu(x)<\infty$ and the cost function $c$ is $c(x,y)=\|x-y\|^2/2$, the next two statements are equivalent:
	\begin{itemize}
		\item[1)] For every $\nu\in\PM(\R^n)$ with $\int|x|^2\ d\nu(x)<\infty$, the optimal plan $\gamma$ between $\mu$ and $\nu$ is unique and induced by a map $T$, i.e it is $\gamma=(Id,T)_\#\mu$.
		\item[2)] $\mu$ is regular.
	\end{itemize}
   If either 1) or 2) hold, the optimal map $T$ is the gradient of a convex function.
\end{thm}
In fact, the convex function whose gradient is optimal is the $c_+$-transform of the $c$-concave function $\varphi$ for which $supp(\gamma)\subset\partial^{c_+}\varphi$.

As already announced above, a similar statement is true for Riemannian manifolds equipped with the Riemmannian distance.

\begin{dfn}[\emph{\textbf{Riemannian metric distance}}]\label{dfn.Rd}
	Let $(M,h)$ be a connected Riemannian manifold. The following formula determines a metric distance on $M$ and is called \emph{Riemannian (metric) distance} or \emph{geodesic distance}:
	\begin{equation}\nonumber
	d(x,y):=\inf_{\gamma}\int_0^1 h(\dot{\gamma},\dot{\gamma})\ dt,
	\end{equation}	
	for $x,y\in M$, where the infimum is taken over all differentiable curves $\gamma$ for which $\gamma(0)=x$ and $\gamma(1)=y$.
\end{dfn}

\begin{remark}
	The topology induced by $d$ coincides with the original topology on $M$. By the Hopf-Rinow theorem, $(M,d)$ is complete as a metric space if and only if $M$ is geodesically complete, i.e. if for all $x\in M$ every geodesic $\gamma(t)$ starting at $x$ is defined for all $t\in\R$.
\end{remark}

\begin{dfn}[\emph{\textbf{Exponential map}}]
	Let $(M,h)$ be a Riemannian manifold. Let $T_xM$ be the tangent space at $x\in M$ and let $U\subset T_xM$ be a neighborhood of the origin $0\in T_xM$ such that the unique geodesic $\gamma_v(t)$ with starting point $x$, i.e. $\gamma_v(0)=x$, and initial velocity $v\in U$, i.e. $\dot{\gamma}(0)=v$, is well-defined at $t=1$. Then the \emph{exponential map} $\exp_x$ is defined in the following way.
	\begin{eqnarray}\nonumber
	\exp_x: U &\longrightarrow& M \\\nonumber
	\ \ \ v\   &\longmapsto & \gamma_v(1).
	\end{eqnarray}
\end{dfn}
According to the Hopf-Rinow theorem, $\exp_x$ can be defined on all of $T_xM$ if and only if $(M,d)$ is a complete metric space.

There is a generalization of regular measures on differentiable manifolds $M$, which we will need:

\begin{dfn}[\emph{\textbf{Regular measure on $M$}}]
	A measure $\mu\in\PM(M)$ is called \emph{regular}, if it assigns no mass to the set of non-differentiability of any semiconvex function.
\end{dfn}
Again, in particular measures which are absolutely continuous with respect to the volume measure are regular. 

We can now cite a variant of McCann's theorem.

\begin{thm}[\emph{\textbf{Existence of optimal maps on manifolds}}]\label{thm.mapmc}
	Let $\mu\in\PM(M)$ be a probability measure on a Riemannian manifold $M$ which is smooth, compact and without boundary. Let further $c=d^2/2$ be the cost function, $d$ the Riemannian metric distance. Then the following two statements are equivalent: 
	\begin{itemize}
		\item[1)] The optimal transport plan between $\mu$ and any other measure $\nu\in\PM(M)$ is unique and induced by a map $T$.
		\item[2)] $\mu$ is regular.
	\end{itemize}
   In these cases, the optimal map $T$ is of the form $T(x)=\exp_x\left(-\nabla\varphi(x)\right)$, where $\varphi:M\rightarrow\R$ is a $c$-concave function.
\end{thm}
Here again, the $c$-concave function $\varphi$ is the one on whose $c$-superdifferential the optimal plan $\gamma$ is concentrated on.

%------------------------------------
\subsection{Wasserstein spaces $W_p(X)$}
%------------------------------------
From now on, we denote the set of probability measures which have finite $p$-th moment by $\PM_p(X)$:
$$\PM_p(X):=\{\mu\in\PM(X)\ \mid \int_X d^p(x_0,x)\ d\mu(x)<\infty,\ x_0\in X\},$$ where $p\in[1,\infty).$
It is sufficient to demand the finiteness of the integral only for one $x_0\in X$. Together with the triangle- and the Minkowski inequality it follows that finiteness holds for every $x_0\in X$.

Given a cost function $c$, to every two probability measures $\mu$ and $\nu$ one can assign to them the number $C(\mu,\nu)$, see formula \eqref{eq.kant}, being the optimal total cost for transporting one measure onto the other.
One might want to think that at least for $X=Y$ this number encodes some kind of distance between $\mu$ and $\nu$. Unfortunately, in general $C(\cdot,\cdot)$ does not satisfy the metric distance axioms. But in case the cost function is a power $p$ of $d$, $C$ is indeed a metric, if restricted to $\PM_p(X)$.

\begin{thm}
	Let $(X,d)$ be a Polish metric space and $p\in[0,\infty)$, then
	\begin{eqnarray*}\nonumber
		W_p: \PM_p(X)\times\PM_p(X)&\rightarrow& X\\\nonumber
		(\mu,\nu) &\mapsto& \left(\inf_{\gamma\in Adm(\mu,\nu)}\int_{X\times X}d^p(x,y)\ d\gamma(x,y)\right)^{1/p}
	\end{eqnarray*}
	is a metric distance.
\end{thm}
\begin{dfn}[\emph{\textbf{Wasserstein distances and Wasserstein spaces}}]\label{df.Wasdist}
	The metric $W_p$ is called \emph{$p$-th Wasserstein distance}, or Wasserstein distance of order $p$. The tuple $(\PM_p(X),W_p)$ is called \emph{Wasserstein space} and is denoted by the symbol $W_p(X)$. Instead of $W_2(X)$ we will often just write $W(X)$.
\end{dfn}

\begin{remark}
	The map $X\rightarrow\PM(X),\ x\mapsto\delta_x$ is an isometric embedding of the underlying Polish space $X$ into the Wasserstein space on $X$, since $W_p(\delta_x,\delta_y)=d(x,y)$ for every $p\in[1,\infty)$. We note furhter that the $p$-th moment of $\mu$ is nothing but the $p$-th Wasserstein distance of $\mu$ to $\delta_{x_0}$ to the power of $p$: $\int_Xd^p(x,x_0)\ d\mu(x)=W_p^p(\mu,\delta_{x_0})$.
\end{remark}
\begin{example}
	Let $\mu=\sum_{i=1}^na_i\delta_{x_i}$ and $\nu=\delta_y$ then $$W_p^p(\mu,\nu)=\sum_{i=1}^na_id^p(x_i,y).$$
\end{example}

It is important to know that the Wasserstein distance $W_p$ metrizes the weak convergence in $\PM_p(X)$. This means that the weak convergence of $(\mu_k)_{k\in\N}$ to $\mu$ in $\PM_p(X)$ is equivalent to $W_p(\mu_k,\mu)\ra 0$. This is a useful property, but not unique to the Wasserstein distances. As a reminder, we give the definition of weak convergence.

\begin{dfn}[\emph{\textbf{Weak convergence in $\PM(X)$}}]
	A sequence $(\mu_k)_{k\in\N}\subset\PM(X)$ is said to \emph{converge weakly} to $\mu\in\PM(X)$ if and only if $\int\varphi d\mu_k\ra\int\varphi d\mu$ for any bounded continuous function $\varphi$ on $X$. This is denoted by $\mu_k\rightharpoondown\mu$.
	A sequence $(\mu_k)_{k\in\N}\subset\PM_p(X)$ is said to converge weakly to $\mu\in\PM_p(X)$ if and only if for $x_0\in X$ it is:
	\begin{itemize}
		\item[1)] $\mu_k\rightharpoondown\mu$ and
		\item[2)] $\int d(x_0,x)^p d\mu_k(x)\ra\int d(x_0,x)^p d\mu(x)$.
	\end{itemize}
	This is denoted by $\mu_k\rightharpoonup\mu$.
\end{dfn}
We further list some more important properties of $W_p(X)$.
\begin{thm}[\emph{\textbf{Some properties of $W_p(X)$}}]\label{thm.prop}
	\begin{itemize}
		\item [1)] $W_p(X)$ is compact, in case $X$ is. On the other hand one can show that whenever $X$ is unbounded, then $W_p(X)$ cannot be locally compact.
		\item [2)] $W_p$ is continuous on $\PM_p(X)\times \PM_p(X)$. That means, if $\mu_k\rightharpoonup\mu$ and $\nu_k\rightharpoonup\nu$, then $W_p(\mu_k,\nu_k)\ra W_p(\mu,\nu)$.
		\item [3)] $W_p(X)$ is complete and separable.
	\end{itemize}
\end{thm}
%
%-------------------------------------------------------------------
\subsection{Wasserstein geodesics and displacement interpolation}
%-------------------------------------------------------------------
%
Let us now have a look at the geodesic structure of Wasserstein space. 
\begin{dfn}[\emph{\textbf{Constant speed geodesic}}]\label{df.geod}
	A curve $(\gamma_t)_{t\in [0,1]}$, $\gamma_0\neq\gamma_1$, in a metric space $(X,d)$ is called a \emph{constant speed geodesic} or \emph{metric geodesic} in case that
	\begin{equation}\label{eq.geod}
	d(\gamma_t,\gamma_s)=|t-s|d(\gamma_0,\gamma_1)\ \ \forall t,s\in [0,1]. 
	\end{equation}
\end{dfn}

In Section \ref{sec.dyn} we will see that constant speed geodesics actually have a constant metric derivative.

We will often abbreviate curves $(\gamma_t)_{t\in [0,1]}$ by omitting the brackets and the interval of definition, i.e we will often just write $\gamma_t$ instead. One can show that a curve $(\gamma_t)_{t\in [0,1]}$, $\gamma_0\neq\gamma_1$, is a geodesic if and only if $d(\gamma_t,\gamma_s)\leq |t-s|d(\gamma_0,\gamma_1)\ \ \forall t,s\in [0,1].$ In particular, this implies that a given geodesic never crosses a point twice. 
We further have the following corollary.

\begin{corollary}
	If a curve $(\gamma_t)_{t\in [0,1]}$ is a constant speed geodesic, then for every $s<u<t$ it is
	\begin{equation}\label{eq.shortgeod}
	d(\gamma_s,\gamma_t)=d(\gamma_s,\gamma_u)+d(\gamma_u,\gamma_t).
    \end{equation}
\end{corollary}

\noindent Thus, one can picture $\gamma_t$ to be the shortest path between its endpoints. However, the converse implication is not true.

\begin{remark}
	Since $x\mapsto \delta_x$ is an isometry, for every constant speed geodesic $\gamma(t)$ in $X$, $\delta_{\gamma(t)}$ is a constant speed geodesic in $W_p(X)$.
\end{remark}
\begin{dfn}[\emph{\textbf{Geodesic space}}]
	A metric space $(X,d)$ is called \emph{geodesic} if for every $x,y\in X$ there exists a constant speed geodesic $(\gamma_t)_{t\in [0,1]}$ with $\gamma_0=x$ and $\gamma_1=y$. 
\end{dfn}
One can prove the following important set of properties.
\begin{thm}[\emph{\textbf{Wasserstein geodesics}}]\label{thm.geod}
	\begin{itemize}
		\item Whenever $(X,d)$ is geodesic, $W_2(X)$ is geodesic as well. 
		\item If $X$  is a Hilbert space, $\mu_t$ is a constant speed geodesic in $W_2(X)$ if and only if there exists an optimal transport plan $\gamma\in Adm(\mu_0,\mu_1)$ such that $$\mu_t=((1-t)\pi^1+t\pi^2)_\#\gamma.$$
		\item If further $\gamma$ is induced by a map $T$, this latter formula simplifies to $$\mu_t=((1-t)Id+tT)_\#\mu_0$$
	\end{itemize}
\end{thm}
\begin{remark}
	Curves of the form $\mu_t=((1-t)\pi^1+t\pi^2)_\#\gamma$ are called \emph{displacement interpolation} from $\pi^1_\#\gamma$ to $\pi^2_\#\gamma$ through $\gamma$. The previous theorem thus tells that within the Wasserstein structure, it is more natural to interpolate measures on the level of measurable sets than on the level of measures itself: The curve $\mu_t=(1-t)\mu_0+t\mu_1$, where interpolation is done by just shifting mass from one measure to the other, has infinite length in Wasserstein space and is, although being continuous, not absolutely continuous. In $((1-t)\pi^1+t\pi^2)_\#\gamma$ the prescription is that a set $A$ has the amount of mass that $\gamma$ is assigning to the set of all those points $(x,y)$ with interpolation $(1-t)x+ty$ being an element of $A$.
\end{remark}

There is an even richer geodesic structure, in case $X$ is a Riemannian manifold.
\begin{thm}[\emph{\textbf{Wasserstein geodesics on manifolds}}]\label{thm.mf}
	 Let $M$ be a smooth Riemannian manifold
	\begin{itemize}
		\item $\mu_t$ is a constant speed geodesic in $W_2(M)$ if and only if there exists a $\gamma\in\PM_2(TM)$ such that $\int|v|^2\ d\gamma(x,v)=W_2^2(\mu_0,\mu_1)$ and $(Exp(t))_\#\gamma=\mu_t$. Here, $Exp(t):TM\rightarrow M,\ (x,v)\mapsto exp_x(tv)$.
		\item The joining constant speed geodesic of two measures in $W_2(M)$ is unique, provided one of the measures is absolutely continuous with respect to the volume measure. 
		\item If $\mu_t$ is a constant speed geodesic in $W_2(M)$, then for every $t\in(0,1)$ and $s\in[0,1]$ there exists only one optimal transport plan from $\mu_t$ to $\mu_s$ and this plan is induced by a map which is locally Lipschitz.
	\end{itemize}
\end{thm}

We want to introduce one further notion that we are going to need later.

\begin{dfn}[\emph{\textbf{Non branching space}}]
	A metric space $(X,d)$ is called \emph{non branching}, if the following map is injective for every $t\in(0,1)$
	\begin{eqnarray}\nonumber
	Geod(X) &\longrightarrow& X\times X\\\nonumber
	\ \gamma\ &\longmapsto& (\gamma_0,\gamma_t).
	\end{eqnarray}
\end{dfn}

One can show that if $(X,d)$ is a complete, separable, locally compact and non branching geodesic space, then also $W_p(X)$ is non branching for $p\in(0,\infty)$. The converse is also true: If $W_p(X)$ is non branching, then so is $X$. A proof of this can be found in \cite{Vilbig} (Corollary 7.32). According to \cite{User}, the local compactness condition on $X$ is, however, not necessary.
%
%------------------------------------------------------------------
\subsection{Dynamical equations on $W_2(M)$}\label{sec.dyn}
%------------------------------------------------------------------
%
In what follows, we will only be concerned with $W_2(M)$, where $(M,h)$ is a smooth, connected and complete Riemannian manifold with Riemannian metric tensor $h$.
Furthermore, we equip the set of measurable sections in $TM$, which we will denote by $\Gamma(TM)$, with an $L^2$-topology.\footnote{A section $v:M\rightarrow TM$ is measurable whenever all components in every chart are measurable. Equivalently, if and only if $v$ itself is measurable with respect to the Borel $\sigma$-algebras given on $M$ and $TM$.}  That means, for $v\in \Gamma(TM)$ we define
\begin{equation*}
	\|v\|_{L^2(\mu)}:=\sqrt{\int_M h(v,v)\ d\mu}
\end{equation*}
and  \begin{equation*}
	L^2(M,\mu):=\bigslant{\{v\in \Gamma(TM)\mid \|v\|_{L^2(\mu)} <\infty\}}{\sim}.
\end{equation*}
Here, two vector fields are considered to be equivalent in case they differ only on a set of $\mu$-measure zero. $L^2(M,\mu)$ is a Hilbert space with the canonical scalar product. We will often write $L^2(\mu)$ if it is clear to which manifold $M$ is referred to.

The goal now is to identify a differentiable structure on $W_2(M)$. Unfortunately, there is no smooth structure in the traditional sense, e.g. in the sense of \cite{Kriegl_1997} where infinite dimensional manifolds are modeled on convenient vector spaces. So one has to try to find structures that resemble formally what one requires of a manifold structure.
The differentiable structure on $W_2(M)$ that will be defined in the end thus consists of ad hoc definitions, accurately tailored to Optimal transport and the Wasserstein metric structure, which mimic conventional differentiable and Riemannian behavior. This could be seen as a misfeature of the theory, but this ad hoc calculus yields powerful tools to perform calculations and, which is particularly important for us, provides a very natural language to capture characteristics of physical theories. 

The basic idea of a tangent vector at a given point is that it indicates the direction a (differentiable) curve will be going infinitesimally from that point. Then traditionally the set of all such vectors which can be found to be tangential to some curve at a given fixed point are collected in the tangent space at that point. 
In our situation, we could enlarge $\PM_2(M)$ to the Banach space of signed measures where one could expect the tangent space at point $\mu$ to be filled with all the Radon measures with zero integral and nonnegative outside $supp(\mu)$. 
However, we would like to take a different approach here which is more adapted to the geometric structure given by the Wasserstein distance. Unfortunately, on $W_2(M)$ there is no notion of smooth curves - but there is a notion of metric geodesics. In case the transport plan for the optimal transport between two measures is induced by a map $T$, the interpolating geodesic on Hilbert spaces (see Theorem \ref{thm.geod}) can be written as $\mu_t=((1-t)Id+tT)_\#\mu_0$, thus being of the form $\mu_t=F_{t\#}\mu_0$. Generally, on Riemannian manifolds optimal transport between $\mu_0$ and $\mu_t$ can be achieved by $\mu_t=F_{t\#}\mu_0$, $F_t=\exp(t\nabla\varphi)$( see e.g \cite{Vilbig}, Chapter 12)\footnote{This formula in particular dispays nicely that optimal transport happens along geodesics of the underlying metric space.}. In these cases, $F_t$ is injective and locally Lipschitz for $0<t<1$ (\cite{Vilsmall}, Subsubsection 5.4.1).
It is known from the theory of characteristics for partial differential equations that curves of this kind solve the weak continuity equation, together with the vector field to which integral lines $F_t$ corresponds. 
\begin{dfn}[\emph{\textbf{Continuity equation}}]\label{dfn.conti}
	Given a family of vector fields $(v_t)_{t\in [0,T]}$, a curve $\mu_t:[0,T]\rightarrow W_2(M)$ is said to solve the \emph{continuity equation}
	\begin{equation}\label{eq.cont}
	\partial_t\mu_t+\nabla\cdot(v_t\mu_t)=0
	\end{equation}
	in the weak sense, in case
	\begin{equation}\label{eq.wcont}
	\int_0^T\int_M\left(\frac{\partial}{\partial t}\varphi(x,t)+h(\nabla\varphi(x,t),v_t(x))\right)\ d\mu_t(x)dt =0
	\end{equation}
	holds true for all $\varphi\in C_c^\infty\left((0,T)\times M\right)$.
\end{dfn}
\begin{thm}\label{thm.conti}
	Let $(F_t)_{t\in[0,T)}$ be a family of maps on $M$ such that $F_{t}:M\rightarrow M$ is a bijection for every $t\in[0,T)$, $F_0=Id$ and both $(t,x)\mapsto F_t(x)$ and $(t,x)\mapsto F_t^{-1}(x)$ are locally Lipschitz on $[0,T)\times M$. Let further $v_t(x)$ be a family of velocity fields on $M$ such that its integral lines correspond to the trajectories $F_t$, and $\mu$ be a probability measure. Then $\mu_t=F_{t\#}\mu$ is the unique weak solution in $\mathcal{C}\left([0,T),\PM(M)\right)$ of $\frac{d}{dt}\mu_t+\nabla\cdot(v_t\mu_t)=0$ with initial condition $\mu_0=\mu$. Here, $\PM(M)$ is equipped with the weak topology.
\end{thm}
%
	% A uniformly Lipschitz vector field generates a flow which is (globally) Lipschitz.
%
	Theorem \ref{thm.conti} is taken from \cite{Vilsmall} where it is labeled as Theorem 5.34. % p.167.

The question now is, whether one can characterize the class of curves on $W_2(M)$ that admit a velocity in the manner of \ref{dfn.conti}. A satisfying answer is given by Theorem \ref{them.ac}, taken from \cite{User}. 
\begin{dfn}[\emph{\textbf{Absolutely continuous curve}}]
Let $(E,d)$ be an arbitrary metric space and $I$ an interval in $\R$. A function $\gamma:I\rightarrow E$ is called \emph{absolutely continuous, a.c.,} if there exists a function $f\in L^1(I)$ such that 
\begin{equation}\label{eq.ac}
d(\gamma(t),\gamma(s))\leq\int_t^s f(r)dr,\ \ \ \forall s,t\in I, t\leq s.
\end{equation}
\end{dfn}
\begin{dfn}[\emph{\textbf{Metric derivative}}] \label{df.md}
The \emph{metric derivative} $|\dot{\gamma}|(t)$ of a curve $\gamma:[0,1]\rightarrow E$ at $t\in (0,1) $ is given as the limit
\begin{equation}
|\dot{\gamma}|(t)=\lim_{h\rightarrow 0}\frac{d(\gamma(t+h),\gamma(t))}{|h|}.
\end{equation}
\end{dfn}
It is known that for absolutely continuous curves $\gamma$, the metric derivative exists for a.e. $t$. It is an element of $L^1(0,1)$ and, up to sets of zero Lebesgue-measure, the minimal function satisfying equation \eqref{eq.ac} for $\gamma$. In this sense absolutely continuous functions enable a generalization of the fundamental theorem of calculus to arbitrary metric spaces.
\begin{example}
	\begin{enumerate}
		\item Every metric geodesic is absolutely continuous and $|\dot{\gamma}|(t)=d(\gamma(0),\gamma(1))$.
		\item Let $E=\R^n$ with the distance induced by the Euclidean norm $\|\cdot\|$, then $|\dot{\gamma}|(t)=\|\frac{d\gamma}{dt}(t)\|$ at any point $t\in(a,b)$ where $\gamma$ is differentiable.
		\item In particular, every curve of Dirac measures $\delta_{\gamma(t)}$ in Wasserstein space is a.c. if and only if $\gamma(t)$ is a.c., in that case their respective metric derivatives coincide.
	\end{enumerate} 
\end{example}
\begin{thm}[\emph{\textbf{Differential characterization of a.c. curves}}]\label{them.ac}
Let $\mu_t:[0,1]\rightarrow W_2(M)$ be an a.c. curve. Then there exists a Borel family of vector fields $(v_t)_{t\in[0,1]}$ on $M$ such that the continuity equation \eqref{eq.wcont} holds and 
\begin{equation*}
\|v_t\|_{L^2(\mu_t)}\leq |\dot{\mu_t}| \text{  for a.e. } t\in(0,1).
\end{equation*}
Conversely, if a curve $\mu_t:[0,1]\rightarrow W(M)$ is such that there exists a Borel family of vector fields $(v_t)_{t\in[0,1]}$ with $\|v_t\|_{L^2(\mu_t)}\in L^1(0,1)$, together with which it satisfies \eqref{eq.wcont}, then there exists an a.c. curve $\tilde{\mu}_t$ being equal to $\mu_t$ for a.e. $t$ and satisfying
\begin{equation*}
|\dot{\tilde{\mu}}_t|\leq\|v_t\|_{L^2(\tilde{\mu}_t)} \text{  for a.e. } t\in(0,1).
\end{equation*} 
\qed
\end{thm}
It can be shown, see \cite{Gigli}, Theorem 1.31, that for metric geodesics, the velocity vector field is
well defined for every $t\in(0, 1)$, not just for a.e. $t$.

In the following we call a pair $(\mu_t,v_t)$ which together solves the continuity equation an \emph{a.c. couple}. We further want to call $v_t$ an \emph{accompanying vector field} for $\mu_t$.

Starting from geodesics we have arrived at the larger class of absolutely continuous curves on $W_2(M)$ and we would like to think about the vector fields $v_t$ as being tangential to $\mu_t$. 

In this treatise, the physical context in which we will encounter the continuity equation again is quantum physics (see Section \ref{sec.hyquant}). The conserved quantity there being $|\psi|^2$, the squared norm of the wave function interpreted as probability density, and $j=|\psi|^2\nabla S$, where $\psi=|\psi| e^{-i S}$. Whereas in quantum physics the continuity equation subjecting $|\psi|^2$ is treated within standard calculus, we are going to interpret it in the weak sense which enables us to use the Wasserstein geometry formalism to express quantum mechanical features. 

After having found a dynamical equation governing displacement interpolation, one can also ask about the dynamics governed by the family of vector fields $v_t$ which corresponds to the trajectories of transportation. The answer is found in the pressureless Euler equation $\frac{dv}{dt}+\nabla_v v=0$, see \cite{Vilsmall}. Physically, it describes the velocity of particles that travel along geodesics without mutual interaction. In the case which is of most interest to us, namely $v=\nabla S(x,t)$, this reduces to a Hamilton-Jacobi equation 
\begin{equation}\label{eq.ham}
\frac{\partial S}{\partial t}+\frac{\|\nabla S\|^2}{2}=0.
\end{equation} 
We will reencounter a very similar equation again in section \ref{ch.quW} when dealing with the Madelung equations.
The flow map of a smooth solution to the system consisting of the continuity equation and equation \eqref{eq.ham} determines optimal transportation for the cost $d^2$ in case there exists a function $u(y)$ such that $-S(t=0,x)=\inf_{y\in M}\frac{d^2(x,y)}{2}-u(y)$ (i.e. if $-S(t=0,x)$ is $\frac{d^2}{2}$-concave).
Equation \eqref{eq.ham} can be solved using the Hopf-Lax formula 
\begin{equation}\nonumber
S(t,x)=\inf_{y\in M}\left(S(t=0,y)+\frac{d^2(x,y)}{2t}\right).
\end{equation}
A rigorous treatment of this aspect can be found in \cite{Vilsmall}, Subsection 5.4.9.
%
%-----------------------------------------------------------------------
\subsection{Induced differentiable structure on $W_2(M)$}\label{sec.diffstr}
%------------------------------------------------------------------------
%
In this section we want to introduce the notion for tangent spaces that is being used at elements of $W_2(M)$.

As we have seen in Theorem \ref{them.ac}, every absolutely continuous curve in $W_2(M)$ admits an $L^1(dt)$ family of $L^2(\mu_t)$-vector fields together with which the continuity equation is satisfied.  We want to think of this family of vector fields $v_t$ as being tangential to the curve. However, $v_t$ is not unique, there are many vector fields which allow for the same motion of the density: Adding another family $w_t$ with the ($t$-independent) property $\nabla(w_t \mu_t)=0$ to $v_t$  does not alter the equation. Luckily, Theorem \ref{them.ac} provides a natural criterion to choose a unique element among the $v_t's$. According to this theorem, there is at least one family $v_t$ such that $|\dot{\mu}_t|=\|v_t\|_{L^2(\mu_t)}$ for almost all $t$, i.e. that is of minimal norm for almost all $t$. Linearity of \eqref{eq.ac} with respect to $v_t$ and the strict convexity of the $L^2$-norms ensure the uniqueness of this choice, up to sets of zero measure with respect to $t$. We want to call such a couple $(\mu_t,v_t)$, where $v_t$ is the unique minimal tangent family for an a.c. curve $\mu_t$, a \emph{tangent couple}. 

It now seems reasonable to define the tangent space at point $\mu$ as the set of $v\in L^2(M,\mu)$ with $\|v\|_\mu\leq\|v+w\|_\mu$ for all $w\in L^2(M,\mu)$ such that $\nabla(w\mu)=0$. This condition for $v\in L^2(M,\mu)$, however, is equivalent to saying that $\int_M h(v,w)\ d\mu=0$ for all $w\in L^2(M,\mu)$ with $\nabla(w\mu)=0$. This in turn is equivalent to the following, which we will take as the definition of the tangent space.

\begin{dfn}[\emph{\textbf{Tangent space $T_\mu W(M)$}}] \label{dfn.tspace}
	The \emph{tangent space} $T_\mu W(M)$ at point $\mu\in W(M)$ is defined as
\begin{equation}\label{eq.tangentspace}
T_\mu W(M):=\overline{\{\nabla\varphi\mid \varphi\in \mathcal{C}^\infty_c(M)\}}^{L^2(M,\mu)}\subset L^2(M,\mu).
\end{equation}	
\end{dfn}

It is not difficult to see that $\text{dim }T_{\delta} W(M)= \text{dim }M$, whereas in most of the cases $\text{dim }T_\mu W(M)=\infty$. We also give the definition of the normal space:
\begin{eqnarray}\nonumber
T^\perp_\mu W(M) & := & \{w\in L^2(M,\mu)\mid \int h(w,v)\ d\mu=0,\ \forall v\in T_\mu W(M)\}	\\\nonumber
& = & \{w\in L^2(M,\mu)\mid \nabla(w\mu)=0\}.
\end{eqnarray}

One can show that if $(\mu_t,v_t)$ is an a.c. couple such that $\|v_t\|_{L^2(\mu_t)}\in L^1(0,1)$, then $(\mu_t, v_t)$ is a tangent couple if and only if $v_t\in T_{\mu_t}W(M)$ for almost every $t\in(0,1)$.

In Section \ref{sec.dyn} we have seen that the velocity $v_t$ field along a geodesic $\mu_t$ comes from time dependent optimal transport (see Proposition 1.32 in \cite{Gigli} for more details). But we can see more generally that also tangent vector fields of a.c. curves can be obtained from optimal transport (\cite{User}, Proposition 2.32):
Given a tangent couple $(\mu_t,v_t)$ in $W(\R^n)$ such that $\mu_t$ is regular for every $t$ and let $T^s_t$ be the optimal transport map from $\mu_s$ to $\mu_t$. Then for a.e. $t\in[0,1]$ it is, with respect to the limit in $L^2(\mu_t)$,
\begin{equation}\label{eq.limit}
v_t=\lim_{s\rightarrow t}\frac{T^s_t-Id}{s-t}.
\end{equation}
A similar statement is true also on manifolds and even in case there does not exist an optimal transport map between $\mu$ and $\nu$ (see \cite{Gigli}, Theorem 1.31).

Next to defining tangent spaces meaningfully, there is a notion of (formal) Riemannian structure on $W_2(M)$, which is based on the following formula due to J.-D. Benamou and Y. Brenier (\cite{Benamou1999}). It shows that the Wasserstein distance $W_2$, having been defined through the, static, optimal transport problem, can be recovered by a dynamic formula, reminiscing the length functional on Riemannian manifolds, defining the Riemannian metric distance.
\begin{thm}[\emph{\textbf{Benamou-Brenier formula}}]\label{thm.bb}
Let $\mu,\ \nu \in\PM_2(M)$, then it is 
\begin{equation}\label{eq.bb}
W(\mu,\nu)=\inf_{(\mu_t,v_t)}\int_0^1\|v_t\|_{L^2(\mu_t)}\ dt,
\end{equation}
where the infimum is taken among all a.c. couples $(\mu_t,v_t)$ such that $\mu_0=\mu$ and $\mu_1=\nu$.
\end{thm}
Because of formula (\ref{eq.bb}), one can interpret the expression $\int_0^1\|v_t\|_{L^2(\mu_t)}\ dt$ as the length of a curve $\mu_t$. It can be shown (as stated in \cite{Vilbig}, Chapter 13) that the infimum is achieved if and only if $\mu$ and $\nu$ allow for an optimal plan. The minimizing curve will then be a tangent couple where $\mu_t$ is a geodesic.

%% file: Kapitel/QuantumDynamics.tex
\section{Quantum dynamics on $W_2(M)$}\label{ch.quW}
%---------------------------------------------------
%
In this section, a solution of the free Schr\"odinger equation is investigated with the tools of optimal transport. The section starts with a brief introduction to aspects of wave mechanics that are interesting to us (more can be found e.g. in \cite{ToQ} and \cite{Starkl2001}).

\subsection{Elements of quantum dynamics}\label{sec.hyquant}

\subsubsection*{Wave mechanics and probability interpretation}

The \emph{Schr\"odinger equation} is the basic equation of motion in quantum physics and  was introduced by Erwin Schr\"odinger in 1926 in a series of four papers, all entitled with ``Quantisierung als Eigenwertproblem" (engl. ``Quantization as an eigenvalue problem") (\cite{Schroedinger1926},\cite{Schroedinger1926b},\cite{Schroedinger1926c},\cite{Schroedinger1926d}). And although the mathematical apparatus describing quantum features has developed enormously since, we will concern ourselves only with \emph{wave mechanics}, i.e. ``that portion of quantum theory that is based on the Schr\"odinger wave equation"(\cite{ToQ}).

The Schr\"odinger equation for a particle with mass $m$ in an external potential $V(x,t)$ is given by
\begin{equation}\label{eq.schroe}
i\hbar\frac{\partial }{\partial t}\psi(x,t)=\left(-\frac{\hbar^2}{2m}\Delta +V(x,t)\right)\psi(x,t),
\end{equation}
where $\Delta$ denotes the Laplace operator with respect to the $x$-variable. It is a partial differential equation for functions $\psi$ on $\R^3\times\R$ and its solutions are called \emph{wave functions} and considered to represent the state of the particle. For the major part of the paper it will be assumed that $V=0$, i.e. mainly the case of a free particle is going to be considered.

In 1926 Max Born proposed to interpret $\psi(x,t)$ in a probabilistic way (\cite{Born1926a}): the squared modulus $\rho(x,t):=|\psi(x,t)|^2$ of the wave function should describe the probability density of finding the particle at point $x$ at time $t$. The condition $\int \rho(x,t)\ d\lambda(x)=1$ should thereby ensure that the probability of finding the particle somewhere in space is one. 
Born further argued (in \cite{Born1926b}) that if $\psi$ is written as a linear combination of orthonormal functions $\psi_n$, i.e. $\psi=\sum_n c_n\psi_n$, the squared modulus of the coefficients $|c_n|^2$ can be interpreted as the probabilities for finding the system in the state $\psi_n$. In particular, if $\psi_n$ is an eigenfunction of a self-adjoint operator $A$, $|c_n|^2$ is the probability of measuring the eigenvalue corresponding to $\psi_n$. Along these lines, $\int \bar{\psi}A\psi\ d\lambda$ can be interpreted as the expectation value of the outcome of a measurement of $A$, where $\bar{\psi}$ denotes the complex conjugation of $\psi$.

The probability density $\rho(x,t)$ satisfies the continuity equation, i.e. a local law of conservation of probability, together with the \emph{current density of probability} $j(x,t):=\frac{\hbar}{2mi}\left(\bar{\psi}(x,t)\nabla\psi(x,t)-\nabla\bar{\psi}(x,t)\psi(x,t)\right)$:

\begin{equation}\label{eq.conti2}
\frac{d}{dt}\rho(x,t)+\nabla\cdot j(x,t)=0.
\end{equation}
(Compare Section \ref{sec.dyn}.)

Writing the wave function in polar form, i.e. $\psi(x,t)=R(x,t)e^{iS(r,t)/\hbar}$, where $R(x,t)=|\psi(x,t)|=\sqrt{\rho(x,t)}$ and $S(x,t)$ a real function (the \emph{phase} of the wave function), yields 
\begin{equation}\nonumber
j(x,t)=\rho(x,t)\frac{1}{m}\nabla S(x,t).
\end{equation}
Substituting the polar form in \eqref{eq.schroe}, one obtains, after some calculus, two equations: The continuity equation for $\rho$ and $j=\rho\frac{1}{m}\nabla S$ as in \eqref{eq.conti2} and a modified Hamilton-Jacobi equation with generating function $S$,
\begin{equation}\label{eq.quanthamjac}
\frac{\partial}{\partial t}S(x,t)+\frac{1}{2m}\|\nabla S(x,t)\|^2+ V(x) = \frac{\hbar^2}{2m}\frac{\Delta \sqrt{\rho(x,t)}}{\sqrt{\rho(x,t)}}.
\end{equation}
Equation \eqref{eq.quanthamjac} differs from a Hamilton-Jacobi equation only in the term 
\begin{eqnarray}\nonumber
Q(x,t):=-\frac{\hbar^2}{2m}\frac{\Delta \sqrt{\rho(x,t)}}{\sqrt{\rho(x,t)}},
\end{eqnarray}
which is called the \emph{quantum potential} and which vanishes in the classical limit $\hbar\rightarrow 0$. 

The continuity equation \eqref{eq.conti2} together with the quantum Hamilton-Jacobi equation \eqref{eq.quanthamjac} were first derived from the Schr\"odinger equation by Erwin Made\-lung in 1926 (\cite{Madelung}) and are therefore called \emph{Madelung equations}. It is important to note that, however, not every solution of the Madelung equations yields a solution of the Schr\"odinger equation. To read more about this, we refer, for example, to \cite{Wallstrom1994}.

If we compare equation \eqref{eq.quanthamjac} for $V=0$ with equation \eqref{eq.ham}, we see that the Madelung equations \eqref{eq.conti2} and \eqref{eq.quanthamjac} can be considered as the quantum version of the equations for optimal transport (we will mention this analogy again Section \ref{sec.madelungwasserstein}).

In Section \ref{sec.dyn} we have seen that the continuity equation characterizes absolutely continuous curves in $W_2(M)$ (Theorem \ref{them.ac}). Thus, if it would hold for $d\mu_t:=\rho(x,t)d\lambda(x)$ that $\int \|x\|^2d\mu_t<\infty$, i.e. if $\mu_t\in W(\R^3)$ for all $t$, and $\|\nabla S(x,t)\|_{L^2(\mu_t)}\in L^1(0,\infty)$, then we would know that the curve of probability measures defined by the Schr\"odinger equation is absolutely continuous in $W(\R^3)$. We will elaborate on that below.

\subsubsection*{Motion in expectation and the spreading of wave packets}

The expected motion of quantum particles behaves in a surprisingly classical manner. 
Let $\<x(t)\>$ be the expected position and $\<v(t)\>$ the expected velocity of the quantum particle at time $t$:
\begin{eqnarray}\nonumber
\<x(t)\> &=& \int_{\R^3}x\ d\mu_t,\\\nonumber
\<v(t)\> &=& \frac{1}{m}\int_{\R^3} \nabla S(x,t)\ d\mu_t.
\end{eqnarray}
Then, for a free particle it is $\frac{d}{dt}\<v(t)\>=0$  (so $\<v(t)\>\equiv\<v\>$ from now on) and
\begin{equation}\nonumber
\<x(t)\>\ =\ \<v\>t+\<x(t=0)\>.
\end{equation} 
Thus, interestingly, the expected velocity $\<v(t)\>$ does not depend on time and in particular it holds $\frac{d}{dt}\<x(t)\>=\<v\>$.

More generally, for a particle under the influence of the force $F(x,t)=-\nabla V(x,t)$ it holds
\begin{equation}\label{eq.Ehrenfest}
\frac{d}{dt}\<x(t)\>=\frac{1}{m}\<p(t)\>\ \  \ \text{  and  }\ \ \ \frac{d}{dt}\<p(t)\>=\<F(x,t)\>,
\end{equation}
where $$\<p(t)\> = \frac{\hbar}{i}\int_{\R^3}\bar{\psi}(x,t)\nabla\psi(x,t)\ d\lambda(x)$$ and  $$\<F(x,t)\>=-\int \bar{\psi}(x,t)\left(\nabla V(x,t)\right)\psi(x,t)\ d\lambda(x).$$ 
Note that $\frac{d}{dt}\<p(t)\>=F(\<x(t)\>,t)$ does not hold true, unless the force is linearly dependent on position, as in the harmonic oscillator or a constant force. 

Equations \eqref{eq.Ehrenfest} are called Ehrenfest equations and can be generalized to arbitrary self-adjoint operators.

But not only the time-development of the expectation value of the position of a free particle is interesting, also its variance $\sigma^2_x(t)$ is, where
\begin{equation}\nonumber
\sigma^2_x(t):=\int_{\R^3}\left(x-\<x(t)\>\right)^2\ d\mu_t.
\end{equation}
Defining 
\begin{equation}\nonumber
\sigma^2_v:=\int_{\R^3}\|\left(\frac{\hbar}{im}\nabla-\<v\>\right)\psi(x,t)\|^2\ d\lambda(x),
\end{equation}
which does not depend on time, the following formula holds true: 
\begin{equation}\label{eq.spreading}
\sigma^2_x(t)=\sigma_v^2\cdot(t-t_1)^2+\sigma^2_x(t_1).
\end{equation}
Here, $t_1$ is the instant in time at which $\sigma^2_x(t)$ attains its minimum value.
Equation \eqref{eq.spreading} describes the spreading of the wave function of a free particle: starting at $t_1$ the variance $\sigma^2_x(t)$ grows monotonously by the term $\sigma_v^2\cdot(t-t_1)^2$. A consequence of the spreading of the wave packet is a growing decrease of information about the location of the particle.

\subsubsection*{The Gaussian wave packet}

It is possible to give a general solution to the initial value problem of the free Schr\"odinger equation. Is $\psi(x',0)$ the wave function at initial time $t=0$, then 
\begin{equation}\label{eq.solution}
\psi(x,t)=\int_{\R^3} K(x-x',t)\psi(x',0)\ d\lambda(x'), 
\end{equation}
where the integration kernel $K$ is given by
\begin{equation}\nonumber
K(x-x',t)=\left(\frac{m}{i\hbar t}\right)^{3/2}\exp\left(i\frac{m}{2\hbar t}(x-x')^2\right).
\end{equation}
$K$ obeys the initial condition $K(x-x',t=0)=\delta(x-x')$ and is a regular solution of the free Schr\"odinger equation for $t\neq 0$. It can be interpreted as an elementary wave emitted at $t=0$ from the point $x'$. The mathematically simplest solution of the Schr\"odinger equation is a monochromatic plane wave $e^{\frac{i}{\hbar}(px-p^2t/2m)}$, which is however not a physical solution because it is not normalizable.

The simplest initial condition $\psi(x,0)$ for the calculation of the integral in equation \eqref{eq.solution} is a \emph{Gaussian function}. The most general form of this function is
\begin{equation}\label{eq.Gaussgeneral}
\psi(x,0)= N \exp\left(-\frac{1}{2}(x-C)M(x-C)\right),
\end{equation}
where $M\in\C^{3\times 3}$ is a complex $3\times 3$ matrix, $C\in\C ^3$ and $N$ a normalization constant. 

Substituting \eqref{eq.Gaussgeneral} into \eqref{eq.solution} and performing an integration yields 
\begin{equation}\nonumber
\psi(x,t)= N \left(\det\left(\mathds{1}+\frac{i\hbar t}{m}M\right)\right)^{-1/2} \exp\left(-\frac{1}{2}(x-C)\left(M^{-1}+\frac{i\hbar t}{m}\mathds{1}\right)^{-1}(x-C)\right),
\end{equation}
where $\mathds{1}$ is the identity element in $\C^{3\times 3}$. 
The simplest case is the spherical case, where $C=0$ and $M_{ij}=\delta_{ij}l^{-2}$. Then, 
\begin{equation}\label{eq.Gaussfinal}
\psi(x,t)= N \left(1+\frac{i\hbar t}{ml^2}\right)^{-3/2}\exp\left(-\frac{1}{2}\left(l^2+\frac{i\hbar t}{m}\right)^{-1}\|x\|^2\right).
\end{equation}
The probability density in that case is 
\begin{equation}\label{eq.Gaussdensity}
\rho(x,t)=|N|^2\left(1+\frac{\hbar^2t^2}{m^2l^4}\right)^{-3/2}\exp\left(-\left(l^2+\frac{\hbar^2t^2}{m^2l^2}\right)^{-1}\|x\|^2\right).
\end{equation}
%
%---------------------------------------------------------------------------
\subsection{Optimal transport for a solution of the free Schr\"odinger equation}\label{sec.otquant}
%---------------------------------------------------------------------------
%
Let us now investigate solutions of the free Schr\"odinger equation of the form \eqref{eq.Gaussfinal} with the tools of Optimal transport.
Therefore, let $\rho(x,t)$ be as in \eqref{eq.Gaussdensity} and $\mu_t$ be defined by $d\mu_t=\rho(x,t)d\lambda(x)$. We will see that with the coefficient $N^2:=(l^2\pi)^{-3/2}$, $\mu_t$ is a probability measure. The following questions then arise naturally: 

\begin{itemize}
	\item[1)] Is $\mu_t$ an absolutely continuous curve?
	\item[2)] Is the gradient of the phase of $\psi(x,t)$ an element of $T_{\mu_t}W(\R^3)$?
	\item[3)] How does the optimal transport map from $\mu_s$ to $\mu_t$ look like?
	\item[4)] What is the cost for transporting optimally from $\mu_s$ to $\mu_t$? I.e. what is the Wasserstein distance between $\mu_s$ and $\mu_t$?
\end{itemize} 

In this section we will tackle these questions one after the other in the Wasserstein space $W(\R^3)=(\PM_2(\R^3),W(\cdot,\cdot))$, where $$W(\mu,\nu)=\inf_{\pi\in Adm(\mu,\nu)}\int_{\R^3\times \R^3}\|x-y\|^2\ d\pi(x,y),$$ i.e. the cost function is $d^2(x,y)=\|x-y\|^2$.

\begin{lemma} 
	If $N^2=(l^2\pi)^{-3/2}$, $\mu_t$ is a probability measure for every $t\in\R_{>0}$. I.e. 
	$$\mu_t=(l^2\pi)^{-3/2}\left(1+\frac{\hbar^2t^2}{m^2l^4}\right)^{-3/2}\exp\left(-\left(l^2+\frac{\hbar^2t^2}{m^2l^2}\right)^{-1}\|x\|^2\right)\ d\lambda(x)\in \PM(\R^3).$$
\end{lemma}

\begin{proof}

	Define $Q:=\left(1+\frac{\hbar^2t^2}{m^2l^4}\right)^{-3/2}$ and $A:=2\left(l^2+\frac{\hbar^2t^2}{m^2l^2}\right)^{-1}$. Then, according to \cite{Zhang},
	\begin{eqnarray}\nonumber
    & &Q\int_{\R^3}\exp\left(-\frac{1}{2}A\|x\|^2\right)\ d\lambda(x) = Q\sqrt{\frac{(2\pi)^3}{A^3}}\\\nonumber
	&=& \pi^{3/2}\left(1+\frac{\hbar^2t^2}{m^2l^4}\right)^{-3/2}\left(l^2+\frac{\hbar^2t^2}{m^2l^2}\right)^{3/2}\\\nonumber
	&=& \pi^{3/2}\left(\left(\frac{1}{l^2}(l^2+\frac{\hbar^2t^2}{m^2l^2}\right)\right)^{-3/2}\left(l^2+\frac{\hbar^2t^2}{m^2l^2}\right)^{3/2}\\\nonumber
	&=& (l^2\pi)^{3/2}.
	\end{eqnarray}
\end{proof}

\begin{lemma}
$\mu_t\in W(\R^3)$ for every $t\in\R_{\geq0}$.
\end{lemma}

\begin{proof}
	Integrals of the form $\int_{\R^3}\|x\|^2\ \exp(-a\|x\|^2)\ d\lambda(x)$ are finite, so the integral $\int_{\R^3}\|x\|^2\ d\mu_t$ is finite, too.
\end{proof}

Let us continue with the question whether $\mu_t$ is an absolutely continuous curve in $W(\R^3)$. In Section \ref{sec.hyquant}, we have already seen that the continuity equation is satisfied by $(\mu_t,\nabla S(x,t))$, in the strong sense (equation \eqref{eq.conti2}), where $S$ is the phase of the wave function, which means that it also holds in the weak sense. According to Theorem \ref{them.ac}, we have proven absolute continuity of $\mu_t$, as soon as we have shown $\|\nabla S(x,t)\|_{L^2(\mu_t)}\in L^1(0,\infty)$.

\begin{remark}
	The condition in Theorem \ref{them.ac} is $\|v_t\|_{L^2(\mu_t)}\in L^1(0,1)$, since $\mu_t$ was defined for $t\in [0,1]$. However, here we have to adapt to a curve on $[0,\infty)$.
\end{remark}

\begin{lemma}
	The gradient of the phase of the wave function \eqref{eq.Gaussfinal} is given by 
	\begin{equation}\nonumber
	\nabla S(x,t)=\frac{\hbar^2 t}{m}\left(l^4+\frac{\hbar^2 t^2}{m^2}\right)^{-1}\ x.
	\end{equation}
\end{lemma}
Instead of calculating $\nabla S$ directly, after extracting $S$ from $\psi$, we want to use the formula $j(x,t)=\rho(x,t)\frac{1}{m}\nabla S(x,t)$ to find an expression for $\nabla S$.
\begin{proof}
	Let us calculate $j(x,t)=\frac{\hbar}{2mi}\left(\bar{\psi}\nabla\psi-\nabla\bar{\psi}\psi\right)$. We use the following abbreviations: 
	$A:=\left(1+\frac{i\hbar t}{ml^2}\right)^{-3/2}$ and $B:=\left(l^2+\frac{i\hbar t}{m}\right)^{-1}$, so that $\psi(x,t)=NA\exp\left(-\frac{1}{2}B\|x\|^2\right)$. Then 
	\begin{eqnarray}\nonumber
	\nabla\psi(x,t) &=& -BAN\exp\left(-\frac{1}{2}B\|x\|^2\right) x,\\\nonumber
	\nabla\bar{\psi}(x,t) &=& -\bar{B}\bar{A}N\exp\left(-\frac{1}{2}B\|x\|^2\right) x,\\\nonumber
	\bar{\psi}\nabla\psi &=& -B|A|^2N^2\exp\left(-\frac{1}{2}(B+\bar{B})\|x\|^2\right)x\\\nonumber
	\nabla\bar{\psi}\psi &=& -\bar{B}|A|^2N^2\exp\left(-\frac{1}{2}(B+\bar{B})\|x\|^2\right) x.
	\end{eqnarray}
	Therefore, 
	\begin{equation}\nonumber
	\frac{\hbar}{2mi}\left(\bar{\psi}\nabla\psi-\nabla\bar{\psi}\psi\right)=-\frac{\hbar}{2mi}N^2|A|^2(B-\bar{B})\exp\left(-\frac{1}{2}(B+\bar{B})\|x\|^2\right) x.
	\end{equation}
	Furthermore, 
	\begin{eqnarray}\nonumber
	|A|^2 &=&\left(1+\frac{\hbar^2t^2}{m^2l^4}\right)^{-3/2}\\\nonumber
	B+\bar{B} &=& 2l^2\left(l^4+\frac{\hbar^2t^2}{m^2}\right)^{-1}= 2\left(l^2+\frac{\hbar^2t^2}{m^2l^2}\right)^{-1}\\\nonumber
	B-\bar{B} &=& -2 i\frac{\hbar t}{m}\left(l^4+\frac{\hbar^2t^2}{m^2}\right)^{-1}.
	\end{eqnarray}
	So that, because $\rho(x,t)=N^2|A|^2\exp\left(-\frac{1}{2}(B+\bar{B})\|x\|^2\right)$, $$j(x,t)=-\frac{\hbar}{2mi}\rho(x,t)(B-\bar{B}) x.$$
	It follows that
	\begin{eqnarray}\nonumber
	\nabla S(x,t) &=& -\frac{\hbar}{2i}(B-\bar{B})x\\\nonumber
	              &=& \frac{\hbar^2t}{m}\left(l^4+\frac{\hbar^2t^2}{m^2}\right)^{-1}x.
	\end{eqnarray}
\end{proof}

\begin{prop}\label{prop.nablaS}
	\begin{equation}\nonumber
	\|\nabla S(x,t)\|_{L^2(\mu_t)}=\sqrt{\frac{3}{2}}\frac{\hbar^2t}{ml^6}\left(1+\frac{\hbar^2 t^2}{m^2l^4}\right)^{-2}.
	\end{equation}
\end{prop}

\begin{proof}
	Define $C:=\left(\frac{\hbar^2t}{m}\left(l^4+\frac{\hbar^2t^2}{m^2}\right)^{-1}\right)$, $B:=(l^2\pi)^{-3/2}\left(1+\frac{\hbar^2t^2}{m^2l^4}\right)^{-3/2}$ and $A:=2\left(l^2+\frac{\hbar^2t^2}{m^2l^2}\right)^{-1}$. Then, again with \cite{Zhang},
\begin{eqnarray}\nonumber
\|\nabla S(x,t)\|^2_{L^2(\mu_t)} &=& \int_{\R^3}\|\nabla S(x,t)\|^2_{\R^3}\ d\mu_t(x) = C^2 \int_{\R^3} \|x\|^2\ d\mu_t(x)\\\nonumber
                                 &=& C^2 \int_{\R^3} \|x\|^2 B \exp\left(-\frac{1}{2}A\|x\|^2\right)\ d\lambda(x)\\\nonumber
                                 &=& C^2B \sqrt{\frac{(2\pi)^3}{2^3\left(l^2+\frac{\hbar^2t^2}{m^2l^2}\right)^{-3}}}\frac{3}{2}\frac{2}{2}\left(l^2+\frac{\hbar^2t^2}{m^2l^2}\right)\\\nonumber
                                 &=& \frac{3}{2}\left(\frac{\hbar^2t^2}{ml^4}\right)^2\left(1+\frac{\hbar^2t^2}{m^2l^4}\right)^{-7/2}l^{-3}
                                 \left(l^2+\frac{\hbar^2t^2}{m^2l^2}\right)^{-1/2}\\\nonumber
                                 &=& \frac{3}{2l^4}\left(\frac{\hbar^2t^2}{ml^4}\right)^2\left(1+\frac{\hbar^2t^2}{m^2l^4}\right)^{-4}.
\end{eqnarray}	
\end{proof}

\begin{corollary}
	\begin{equation}\nonumber
	\int_0^\infty \|\nabla S(x,t)\|_{L^2(\mu_t)}\ dt = \sqrt{\frac{3}{2}}\frac{m}{2l^2}<\infty.
	\end{equation}
\end{corollary}

\noindent This expression is in particular independent of $\hbar$.

\begin{proof}
	Define $a:=\sqrt{\frac{3}{2}}\frac{\hbar^2}{ml^6}$ and $b:=\frac{\hbar^2}{m^2l^4}$. Then 
	\begin{eqnarray}\nonumber
	\int_0^\infty \|\nabla S(x,t)\|_{L^2(\mu_t)}\ dt &=& \int_0^\infty \frac{at}{\left(1+bt^2\right)^2}\ dt\\\nonumber
	                                                 &=& \frac{a}{2b}.
	\end{eqnarray}
\end{proof}

\begin{corollary}
	The curve $\mu_t$ is absolutely continuous up to a redefinition on a null set in time. \hfill $\square$
\end{corollary}

Let us now approach the question of how the probability distributions of the location of the particle at several times can be transported optimally onto each other. A natural transport map seems already to be given, namely the (time-dependent) flow of $\frac{1}{m}\nabla S$. But is it also optimal? In any case, it is worth finding out, how much it costs when the probabilities are transported with this flow map and whether there is another map that transports at lower cost than the map that is naturally given by the Schr\"odinger equation.

We first recall the definition of the flow of time-dependent vector fields. 
\begin{dfn}
	Let $V(x,t)$ be a time-dependent vector field on $\R^3$, i.e. $V:\R^3\times\R\rightarrow\R^3$. An \emph{integral curve} of $V$ is given by a function $\varphi(x_0,t_0,\cdot):\R\rightarrow\R^3,\ t\mapsto \varphi(x_0,t_0,t)$ which satisfies 
	\begin{equation}\nonumber
	\dot{\varphi}(x_0,t_0,t)=V(\varphi(x_0,t_0,t),t)\ \ \text{ and }\ \ \varphi(x_0,t_0,t_0)=x_0,
	\end{equation}
	where the dot denotes the differentiation with respect to $t$. A \emph{flow map} of $V$ is then given by
	\begin{equation}\nonumber
	F_{s}:\R^3\times\R\longrightarrow\R^3,\ (x,t) \mapsto\ \varphi(x,s,t).
	\end{equation}
	The map $F:(\R^3\times\R)\times\R\rightarrow (\R^3\times\R),\ (x,s,t)\mapsto (\varphi(x,s,t),s+t)$ is a flow map in the time-independent sense, i.e. it satisfies the group law in the last variable.
\end{dfn}
\begin{lemma}\label{eq.flow}
	The (time-dependent) flow of the vector field $\frac{1}{m}\nabla S(x,t)$ is given by
	\begin{equation}\nonumber
	F_s(x,t)=\sqrt{\frac{1+\frac{\hbar^2}{m^2l^4}t^2}{1+\frac{\hbar^2}{m^2l^4}s^2}}\ x.
	\end{equation}
\end{lemma}
\begin{proof}
	Define $V(x,t):=\frac{1}{m}\nabla S(x,t)$. Functions of the form $$\varphi(x_0,t_0,t)=\exp\left(\frac{1}{2}\ln\left(l^4+\frac{\hbar^2t^2}{m^2}\right)+C\right)x$$ are solutions of $\dot{\varphi}(x_0,t_0,t)=V(\varphi(x_0,t_0,t),t)$. Additionally, the initial condition $\varphi(x_0,t_0,t_0)=x_0$ holds, if the constant $C$ is chosen to be $C=-\frac{1}{2}\ln\left(l^4+\frac{\hbar^2t_0^2}{m^2}\right)$. Application of calculation rules for the logarithm yields formula \eqref{eq.flow}.
\end{proof}
\begin{corollary}\label{cor.flow}
	For every pair of times $s$ and $t$, $\mu_t=F_s(\cdot,t)_\#\mu_s$.
\end{corollary}
	We could infer this statement from Theorem \ref{thm.conti}. However, for the sake of directness, we will perform the explicit calculation. For this, we will need the following lemma which we give without proof.
\begin{lemma}\label{lem.imagedensity}
	Let the measure $\mu$ be defined by $d\mu=\rho(x)d\lambda(x)$ and let $T:\R^3\rightarrow\R^3$ be a bijective measurable map. Then, for a measurable set $A$,
	$$T_\#\mu(A)=\int_A (\rho\circ T^{-1})(x)\ dT_\#\lambda(x).$$
	In particular, in case $T\in ISO(\R^3)$, the density of $T_\#\mu$ with respect to the Lebesgue measure is given by $(\rho\circ T^{-1})(x)$.
\end{lemma}	
\begin{proof}[Proof of Corollary \ref{cor.flow}]
	First we see that $$F_s(\cdot,t)_\#\lambda=\left(\frac{1+\frac{\hbar^2}{m^2l^4}t^2}{1+\frac{\hbar^2}{m^2l^4}s^2}\right)^{-3/2}\lambda.$$
	Furthermore, 
	\begin{eqnarray}\nonumber
	& & \rho(F_s^{-1}(x,t),s)\\\nonumber
	 &=& N^2 \left(1+\frac{\hbar^2}{m^2l^4}s^2\right)^{-3/2}\exp\left(-\left(l^2+\frac{\hbar^2s^2}{m^2l^2}\right)^{-1}\frac{1+\frac{\hbar^2s^2}{m^2l^4}}{1+\frac{\hbar^2t^2}{m^2l^4}}\|x\|^2\right)\\\nonumber
	&=& N^2 \left(1+\frac{\hbar^2}{m^2l^4}s^2\right)^{-3/2}\exp\left(-l^{-2}\left(1+\frac{\hbar^2s^2}{m^2l^4}\right)^{-1}\frac{1+\frac{\hbar^2s^2}{m^2l^4}}{1+\frac{\hbar^2t^2}{m^2l^4}}\|x\|^2\right)\\\nonumber
	&=& N^2 \left(1+\frac{\hbar^2}{m^2l^4}s^2\right)^{-3/2}\exp\left(-\left(l^2+\frac{\hbar^2t^2}{m^2l^2}\right)^{-1}\|x\|^2\right)\\\nonumber
	&=& \rho(x,t).
	\end{eqnarray}
	So in total, using Lemma \ref{lem.imagedensity}, $\mu_t=F_s(\cdot,t)_\#\mu_s$.
\end{proof}
\begin{thm}\label{prop.opti}
	The map $F_s(\cdot,t)$ gives the optimal transport from $\mu_s$ to $\mu_t$ for each pair $t,s\in [0,\infty)$.
\end{thm}

\begin{remark}\label{rem.d^2/2}
	In the proof of Theorem \ref{prop.opti} we are going to use the cost function $d^2/2$, as it allows us to use Proposition \ref{prop.later}, which is very convenient. However, $T$ is an optimal map with respect to the cost function $d^2/2$ if and only if it is an optimal map with respect to $d^2$: Let $W^{d^2/2}(\cdot,\cdot)$ be the Wasserstein distance with respect to $d^2/2$ and $W^{d^2}(\cdot,\cdot)$ the one with respect to $d^2$, then if we assume $T$ to be optimal with respect to $d^2/2$,

\begin{eqnarray}\nonumber
W^{d^2/2}(\mu,\nu)^2 &=& \int_X \frac{d^2(x,T(x))}{2}\ d\mu(x)= \inf_{\pi\in Adm(\mu,\nu)}\int_{X\times X} \frac{d^2(x,y)}{2}\ d\pi(x,y) \\\nonumber
                     &=& \frac{1}{2}\inf_{\pi\in Adm(\mu,\nu)}\int_{X\times X}d^2(x,y)\ d\pi(x,y)\\\nonumber
                     &=& \frac{1}{2} W^{d^2}(\mu,\nu)^2.
\end{eqnarray}
So $\int_X d^2(x,T(x))\ d\mu(x)=W^{d^2}(\mu,\nu)^2$ which means that $T$ is optimal with respect to $d^2$, too. Similarly the other way round. In particular, $\mu_t$ is a geodesic with respect to $W^{d^2/2}$ if and only if it is a geodesic with respect to $W^{d^2}$. Whenever we write $W(\cdot,\cdot)$, i.e. whenever we omit the superscript, we refer to the cost function $d^2$.
 \end{remark}

\begin{proof}[Proof of Theorem \ref{prop.opti}]
	Define $C:=\sqrt{\frac{1+\frac{\hbar^2}{m^2l^4}t^2}{1+\frac{\hbar^2}{m^2l^4}s^2}}$ and $\bar{\varphi}(x):=\frac{1}{2}C\|x\|^2$. (In the notation of $C$ and $\varphi$ we neglect the fact that they depend on the parameters $s$ and $t$.) 
	Then $F_s(x,t)=Cx$ and $$F_s(x,t)=\nabla \bar{\varphi}(x).$$
	This, in particular, means that $F_s(x,t)\in\partial^-\bar{\varphi}(x)$, i.e. $F_s(x,t)$ is an element of the subdifferential of $\bar{\varphi}$ at point $x$.
	If we further define $\varphi(x):=\frac{1-C}{2}\|x\|^2$, we see that $\bar{\varphi}(x)=\frac{\|x\|^2}{2}-\varphi(x)$. 
	From Proposition \ref{prop.later} we can conclude that $\varphi$ is $c$-concave and that $F_s(x,t)\in\partial^{c_+}\varphi(x)$. 
	So, if it were that condition \eqref{eq.condition} holds for $\mu_s$ and $\mu_t=F_s(\cdot,t)_\#\mu_s$, we could infer from Remark \ref{rem.later} that $F_s(\cdot,t)$ is optimal from $\mu_s$ to $\mu_t$.
	Regarding condition \eqref{eq.condition}, we first see that  $\frac{\|x-y\|^2}{2}\leq\frac{\|x\|^2}{2}+\frac{\|y\|^2}{2}$. Since integrals of the type $\int_{\R^3}\|x\|^2\exp\left(-B\|x\|^2\right)\ d\lambda(x)$, $B\in\R_{>0}$, are finite, we know that $\frac{\|x\|^2}{2}\in L^1(\mu_{s,t})$.
\end{proof}

The finding of the optimal transport maps enables us to determine whether $\frac{1}{m}\nabla S(x,t)$ is a family of tangent vector fields along $\mu_t$.

\begin{corollary}
	It is $\frac{1}{m}\nabla S(x,t)\in T_{\mu_t}W(\R^3)$ for all $t\in [0,\infty)$.
\end{corollary}

\begin{proof}
	We are going to use formula \eqref{eq.limit}. Since from $\|v_n-v\|^2_{\R^3}\rightarrow 0$ it follows $\|v_n-v\|_{L^2(\mu_t)}\rightarrow 0$, we will only determine the limit in $\R^3$. Let $c:=\frac{\hbar^2}{m^2l^4}$, then
	\begin{equation}\nonumber
	\lim_{s\rightarrow t}\frac{F_t(x,s)-x}{s-t}\ =\ \lim_{s\rightarrow t}\frac{\sqrt{\frac{1+cs^2}{1+ct^2}}\ x-x}{t-s}
	\ =\ \lim_{s\rightarrow t}\frac{\sqrt{\frac{1+cs^2}{1+ct^2}}-1}{t-s}\ x.
	\end{equation}
	Calculating the following limit in $\R$ yields: $$\lim_{s\rightarrow t}\frac{\sqrt{\frac{1+cs^2}{1+ct^2}}-1}{t-s}\ =\ \frac{cs}{cs^2+1}\ =\ \frac{\hbar^2 t}{m}\left(l^4+\frac{\hbar^2 t^2}{m^2}\right)^{-1},$$
	which is precisely the prefactor of $\frac{1}{m}\nabla S(x,t)$. From this we can infer the claimed statement.
\end{proof}

From Theorem \ref{them.ac} we can infer the following corollary.

\begin{corollary}
	The metric derivative $|\dot{\mu}_t|$ of $\mu_t$ equals $\|\nabla S(x,t)\|_{L^2(\mu_{t})}$, i.e. 
	\begin{equation}\nonumber
	|\dot{\mu}_t|=\sqrt{\frac{3}{2}}\frac{\hbar^2t}{ml^6}\left(1+\frac{\hbar^2 t^2}{m^2l^4}\right)^{-2}.
	\end{equation}
	\hfill $\square$
\end{corollary}

\noindent In particular, $\mu_t$ cannot be a constant speed geodesic, as its metric derivative is not constant. However we will see that it is still a geodesic in the sense of formula \eqref{eq.shortgeod}, i.e. in the sense of shortest paths.

We close this section with finally calculating the cost for the optimal transportation from $\mu_s$ to $\mu_t$.

\begin{thm}\label{prop.distance}
	For each $s,t\in[0,\infty)$,
	\begin{equation}\nonumber
	W(\mu_s,\mu_t)=\sqrt{\frac{3}{2}}\ \abs*{\sqrt{1+\frac{\hbar^2 s^2}{m^2 l^4}}-\sqrt{1+\frac{\hbar^2 t^2}{m^2 l^4}}\ }.
	\end{equation}
\end{thm}

\begin{proof}
	Let us again first use the cost function $d^2/2$ and let again $C:=\sqrt{\frac{1+\frac{\hbar^2}{m^2l^4}t^2}{1+\frac{\hbar^2}{m^2l^4}s^2}}$ (dependent on the parameter $s$ and $t$). Then
	\begin{eqnarray}\nonumber
	& & W^{d^2/2}(\mu_s,\mu_t)^2 = \frac{1}{2}\int_{\R^3}\|x-F_s(x,t)\|^2\ d\mu_s(x)\\\nonumber
	& =& \frac{1}{2}\int_{\R^3}\|x-Cx\|^2\ d\mu_s(x)\
	               =\ \frac{1}{2}\int_{\R^3}(1-C)^2\|x\|^2\ d\mu_s(x)\\\nonumber
	               &=& \frac{1}{2}\int_{\R^3}(1-C)^2|N|^2\left(1+\frac{\hbar^2s^2}{m^2l^4}\right)^{-3/2} \exp\left(-\left(l^2+\frac{\hbar^2s^2}{m^2l^2}\right)^{-1}\|x\|^2\right)\ d\lambda(x).
	\end{eqnarray}
	Define now $Q:=(1-C)^2|N|^2\left(1+\frac{\hbar^2s^2}{m^2l^4}\right)^{-3/2}$ and $A:=2 \left(l^2+\frac{\hbar^2s^2}{m^2l^2}\right)^{-1}$.
	Then, with \cite{Zhang},
	\begin{eqnarray}\nonumber
	W^{d^2/2}(\mu_s,\mu_t)^2 &=& \sqrt{\frac{(2\pi)^3}{A^3}}\frac{3}{2}\frac{Q}{A}=\frac{3\sqrt{(2\pi)^3}}{2}\frac{Q}{A^{5/2}}.
	\end{eqnarray}
	Let us first determine the factor $\frac{Q}{A^{5/2}}$ separately.
	\begin{eqnarray}\nonumber
	\frac{Q}{A^{5/2}} &=& \frac{(1-C)^2N^2}{2^{5/2}}\frac{\left(1+\frac{\hbar^2s^2}{m^2l^4}\right)^{-3/2}}{\left(l^2+\frac{\hbar^2s^2}{m^2l^2}\right)^{-5/2}}
	= \frac{(1-C)^2N^2}{2^{5/2}}\frac{\left(l^2+\frac{\hbar^2s^2}{m^2l^2}\right)^{5/2}}{\left(\frac{1}{l^2}\left(l^2+\frac{\hbar^2s^2}{m^2l^2}\right)\right)^{3/2}}\\\nonumber
	&=& \frac{(1-C)^2N^2}{2^{5/2}}l^3\left(l^2+\frac{\hbar^2s^2}{m^2l^2}\right).
	\end{eqnarray}
	Substituting $C$ and $N$ yields
	\begin{eqnarray}\nonumber
	\frac{Q}{A^{5/2}} &=& \frac{l^3(l^2\pi)^{-3/2}}{2^{5/2}} \left(l^2+\frac{\hbar^2s^2}{m^2l^2}\right) \left(1-\sqrt{\frac{1+\frac{\hbar^2t^2}{m^2l^4}}{1+\frac{\hbar^2s^2}{m^2l^4}}}\right)^2\\\nonumber
	&=& \frac{l^5}{(l^2\pi)^{3/2}2^{5/2}}\left(\sqrt{1+\frac{\hbar^2s^2}{m^2l^4}}\right)^2\left(1-\sqrt{\frac{1+\frac{\hbar^2t^2}{m^2l^4}}{1+\frac{\hbar^2s^2}{m^2l^4}}}\right)^2\\\nonumber
	&=& \frac{l^5}{l^3(\pi)^{3/2}2^{5/2}} \left(\sqrt{1+\frac{\hbar^2s^2}{m^2l^4}}-\sqrt{1+\frac{\hbar^2t^2}{m^2l^4}}\right)^2.
	\end{eqnarray}
	So in total, 
	\begin{eqnarray}\nonumber
	\frac{3\sqrt{(2\pi)^3}}{2}\frac{Q}{A^{5/2}} &=& \frac{3\sqrt{(2\pi)^3}}{2}\frac{l^2}{(\pi)^{3/2}2^{5/2}} \left(\sqrt{1+\frac{\hbar^2s^2}{m^2l^4}}-\sqrt{1+\frac{\hbar^2t^2}{m^2l^4}}\right)^2\\\nonumber
	&=& \frac{3}{4}l^2\left(\sqrt{1+\frac{\hbar^2s^2}{m^2l^4}}-\sqrt{1+\frac{\hbar^2t^2}{m^2l^4}}\right)^2.	
	\end{eqnarray}
\end{proof}

\begin{corollary}\label{cor.qugeod}
	For $\mu_t$ as before and $s,t,u\in\R_{\geq0}$, $s<u<t$, we have
	\begin{equation}\nonumber
	W(\mu_s,\mu_t)=W(\mu_s,\mu_u)+W(\mu_u,\mu_t),
	\end{equation}	
	for all $t\in\R_{\geq0}$.
\end{corollary}

\begin{proof}
	\begin{multline}\nonumber
	 W(\mu_s,\mu_u)+W(\mu_u,\mu_t)
		 =  \sqrt{\frac{3}{2}}\ \left(\sqrt{1+\frac{\hbar^2 u^2}{m^2 l^4}}-\sqrt{1+\frac{\hbar^2 s^2}{m^2 l^4}}\right)+\\  \sqrt{\frac{3}{2}}\ \left(\sqrt{1+\frac{\hbar^2 t^2}{m^2 l^4}}-\sqrt{1+\frac{\hbar^2 u^2}{m^2 l^4}}\right)
		                              =W(\mu_s,\mu_t).
	\end{multline}
\end{proof}
%
%------------------------------------------------------------
\subsection{Fisher Information}\label{sec.madelungwasserstein}
%------------------------------------------------------------
%
This section briefly discusses the article \cite{von2012optimal} by M.-K. von Renesse and connects it to our previous analysis. Following up a work by Lott \cite{Lott_2007}, Renesse formally used notions for higher order calculus on $W(\R^3)$ to be able to express the Madelung equations \eqref{eq.conti2} and \eqref{eq.quanthamjac} in the following unified way\footnote{A thorough mathematical treatment of second order analysis on Wasserstein space has been carried out by Gigli in \cite{Gigli}.}: 
\begin{equation}\label{eq.Renesse}
\nabla^W_{v_t}v_t=-\nabla^{L^2(\mu_t)}\left(V(\mu_t)+\frac{\hbar^2}{8}I(\mu_t)\right),
\end{equation}
where $$V(\mu):=\int_{\R^3}V(x)\ d\mu,$$ with $V(x)$ the classical potential from the Schr\"odinger equation \eqref{eq.schroe} and $$I(\mu):=\int_{\R^3}\|\nabla \ln(\rho(x))\|^2\ d\mu,$$ the so called \emph{Fisher information} for $d\mu=\rho(x)d\lambda$.
Furthermore, the symbol $\nabla^W_{v_t}v_t$ means the formal covariant derivative in $W(\R^3)$ of $v_t$ along $v_t$, which we are not going to explain in more detail. Finally, $\nabla^{L^2(\mu)}$ denotes the gradient with respect to the canonical $L^2(\mu)$ scalar product.

The most stunning aspect of this formula is obviously the classical coat of this equation: \eqref{eq.Renesse} has the form of Newton's second law of motion, where the potential of the force is given as the sum of the mean value of the classical potential $V(x)$ and the Fisher Information $I(\mu)$. 
It is also interesting to note that in particular, the Fisher information precisely corresponds to the quantum potential in formula \eqref{eq.quanthamjac} and therefore is the part in the equation that makes the dynamics ``quantum''. 
According to \cite{von2012optimal}, taking the limit $\hbar\rightarrow 0$, solutions of the classical Newtonian equation $\gamma(t)$ solve \eqref{eq.Renesse} after applying the canonical isometric embedding from $\R^3$ to $W(\R^3)$, $\gamma_t\mapsto\delta_{\gamma_t}$. 

To connect this to our previous analysis, we now calculate the value of the Fisher information along the curve $\mu_t$ studied in Section \ref{sec.otquant}.

\begin{corollary}
	Let $\mu_t$ be as in \eqref{eq.Gaussdensity} with $N^2=(l^2\pi)^{-3/2}$. Then for every $t\in\R_{\geq0}$,
	\begin{equation}\nonumber
	I(\mu_t)\ = \frac{6}{l^4}\left(1+\frac{\hbar^2t^2}{m^2l^4}\right)^{-4}\ =\ 4\ \left(\frac{ml^2}{\hbar^2t}\right)^2\ \|\nabla S(x,t)\|^2.
	\end{equation}
	For $t=0$ it is then, $I(\mu_0)=\frac{6}{l^4}$. In particular, $I(\mu_t)$ is monotonously decreasing. 
\end{corollary}

That $I(\mu_t)$ is monotonously decreasing makes sense since $\mu_t$ is the squared norm of a wave packet that is spreading, which means that the information about the location of the particle is decreasing.

Let us also recall that $\|\nabla S(x,t)\|_{L^2(\mu_t)}=|\dot{\mu}_t|$, so that also $I(\mu_t)=4\frac{l^2m}{\hbar^2t}|\dot{\mu}_t|$.

\begin{proof}
	Define $A:=\left(l^2+\frac{\hbar^2t^2}{m^2l^2}\right)^{-1}$, $B:=(l^2\pi)^{-3/2}\left(1+\frac{\hbar^2t^2}{m^2l^4}\right)^{-3/2}$ and $C:=\left(\frac{\hbar^2t}{m}\left(l^4+\frac{\hbar^2t^2}{m^2}\right)^{-1}\right)$. 
	Then $C=\frac{\hbar^2t}{ml^2}A$ and 
	\begin{eqnarray}\nonumber
	I(\mu_t) &=& \int_{\R^3}\|\nabla \ln(\rho(x,t))\|^2\ d\mu(x)\\\nonumber 
	&=& \int_{\R^3}\|\nabla\left(\ln B+\ln\left(\exp\left(-A\|x\|^2\right)\right)\right)\|^2\ d\mu(x) \\\nonumber
	&=& \int_{\R^3}A^2\|\nabla \|x\|^2 \|^2\ d\mu(x) = \int_{\R^3}A^2 \|2x\|^2 = 4A^2\int_{\R^3}\|x\|^2\ d\mu(x) \\\nonumber
	&=& 4\left(\frac{ml^2}{\hbar^2t}\right)^2C^2\int_{\R^3}\|x\|^2\ d\mu(x).
	\end{eqnarray}
	Comparing the last expression with the calculation of the proof of Proposition \ref{prop.nablaS}, we get the claimed formula.
\end{proof}

%% file: Kapitel/DynamicsShapeSpace.tex
\section{Infinitesimal dynamics on Shape space} \label{ch.infSh}
%----------------------------------------------------------------
%
This section aims to combine the analysis of the behaviour of free quantum particles in terms of optimal transport started above with the study of so-called Shape spaces defined in a previous paper \cite{Lessel2025a}. Shape spaces define equivalence classes of measures in which they only differ by their location on space. That is, Shape spaces are thought to define a notion of \emph{shapes} of measures. Subsection \ref{s.shapespaces} recalls the definitions and properties needed here, including a definition of tangent spaces on a specific Shape space. 

The question for the remaining section will then be whether or not it makes sense to describe the evolution of a free quantum particle only in terms of change of shapes. More concretely, we want to capture how much information is lost when projecting a solution $\mu_t$ to its image curve $[\mu_t]$ in Shape space and whether it is possible to find an intrinsic description of this dynamics in Shape space.
\subsection{Shape Spaces $\Sh_p(X)$}\label{s.shapespaces}
%---------------------------------------------
%
Let $(X,d)$ be again a Polish metric space, $ISO(X)$ its isometry group and $p\in\N_{\ge 1}$. Every subgroup $G$ of $ISO(X)$ acts naturally on $\PM_p(X)$ by means of the pushforward of measures: $(g,\mu) \mapsto g\mu:=g_\#\mu$. 

\begin{dfn} \label{dfn.equiv}
	Two elements in $\PM_p(X)$ are defined to be \emph{equivalent} if there exists an element in $G$ which maps them into each other:
	\begin{equation*}
		\mu\sim\nu :\Lra\ \exists\ g\in G: \mu=g\nu.
	\end{equation*}
	The quotient space with respect to this equivalence relation is denoted by $\Sh_p(X):=\bigslant{\PM_p(X)}{\sim}$.
\end{dfn}

\begin{dfn}[\emph{\textbf{Shape distance}}]
	The \emph{p-th Shape distance} $D_p(\mu,\nu)$ of $\mu,\nu\in\PM_p(X)$ is defined as 
	\begin{equation*}
		D_p(\mu,\nu):= \inf_{g\in G}{W_p(g\mu,\nu)}. 
	\end{equation*}
\end{dfn}

One can show \cite{Lessel2025a} that this \emph{Shape distance} is a pseudometric on $\PM_p(X)$ modulo $G$. If $X$ is such that the closed balls are compact and the action of $G$ on $X$ is proper, $D_p(\cdot,\cdot)$ is a metric on that space. These are conditions under which the existence of a minimizer for the Shape distance problem can be proven and which we want to assume from now on. In that case, we can in particular interpret $D$ to be a map on $\Sh_p(X)$.
	\begin{eqnarray}\nonumber
	D_p(\cdot,\cdot):\Sh_p(X)\times \Sh_p(X) &\lra& \R\\ \nonumber
	\left([\mu],[\nu]\right) \;\;\quad &\lmapsto& D_p(\mu,\nu). 
\end{eqnarray}

\begin{dfn}[\emph{\textbf{Shape space}}]
	The space $\Sh_p(X)$, together with $D_p(\cdot,\cdot)$ is called \emph{Shape space}. If $p=2$, we will often just write $\Sh(X)$ instead of $\Sh_2(X)$.
\end{dfn}

It is interesting to note that $(\Sh_p(X),D_p)$ is complete and is, in fact, a Polish space itself. Furthermore, if $(X,d_X)\cong(Y,d_Y)$ then $(W_p(X)/G^X, D^X_p)\cong (W_p(Y)/G^Y, D^Y_p)$ if $G^X$ and $G^Y$ are the full isometry groups. This shows that Shape spaces only depend on the metric structure of the underlying space $X$ and the chosen subgroup of the isometry group (see again \cite{Lessel2025a} for more). 

For the study of curves in Shape space, constant speed geodesics play a crucial role (henceforth, we alway mean constant speed geodesics when speaking about geodesics, unless otherwise stated). One can find conditions under which a geodesic in $W_p(X)$ is also (after projection) a geodesic in $\Sh_p(X)$. However, it is also possible that other continuous curves can be projected on to geodesics in $\Sh_p(X)$. We give without proof (see again \cite{Lessel2025a}) the following statement
 
\begin{prop}
	Let $X$ be such that $W_p(X)$ is a non-branching geodesic space and such that the path-connected component of the identity element $Id_G$ of $G$ is not trivial. Then there exists a continuous curve $\tilde{\mu}_t:[0,1]\rightarrow W_p(X)$ which is not a geodesic but its projection onto Shape space is.
\end{prop}

One way to construct a curve that is projected onto a geodesic but which itself is none is the following.

\begin{lemma}
	Let $X$ be such that $W_p(X)$ is a non-branching geodesic space. In $W_p(X)$ let $\mu_t$ be a geodesic such that there is a $g\in G$ with $g\mu_1\neq\mu_1$. Then, if $F:[0,1]\rightarrow[0,1]$ is a continuous function with $F(t)=1$ for all $t\leq t_0<1$ but $F(t)\neq1$ for all $t>t_0$, the mixing $\tilde{\mu}_t:=\left(F(t)+(1-F(t)g)\right)_\#\mu_t$ is not a geodesic, but its projection onto Shape space is.
\end{lemma}

Finally, one can show \cite{Lessel2025a} that if $W_p(X)$ is geodesic, then $\Sh_p(X)$ is geodesic, too. (And we know from existing literature that $W_2(X)$ is geodesic whenever $(X,d)$ is geodesic.)

Important special cases of Shape spaces are those in which $X$ is a Riemannian manifold $(M,h)$. These are separable and locally compact metric spaces by definition and the metric topology coincides with the manifold topology. However, in general Riemannian manifolds are not complete, i.e. not Polish, but for our purposes it is of course necessary to require this from now on. One can prove the following important statement (\cite{Lessel2025a}).
\begin{thm}
	In case the Polish space $X$ is a complete connected Riemannian manifold and its isometry group is equipped with the compact-open topology, a minimizer for the Shape distance-problem exists and $D_p(\cdot,\cdot)$ is a metric distance on $\Sh_p(X)$.
\end{thm}

The remainder of this subsection gives a definition for tangent spaces on $\Sh(M)$, as proposed in \cite{Lessel2025a}. Let thus $(M,h)$ be a smooth, connected, complete Riemannian manifold. The tangent space construction leans on the finite dimensional situation in which the following holds true:
In case a finite dimensional Lie group $G$ acts smoothly, freely and properly on a finite dimensional smooth manifold $M$, the quotient space $M/G$ is a topological manifold of dimension dim $M/G=$ dim $M\ -$ dim $G$. The quotient space furthermore has a unique smooth structure with respect to which the quotient map $\pi:M\rightarrow M/G$ is a smooth submersion (see \cite{Lee2001}). 
The orbits $G.x$, $x\in M$, then, can proven to be embedded submanifolds of $M$ and the kernel of the linear and surjective map $d\pi_x:T_xM\rightarrow T_{\pi(x)}\left(M/G\right)$ is precisely $T_x(G.x)$ (for these statements see again \cite{Lee2001}, Lemma 5.29 and Theorem 7.5). This yields the following isomorphism of linear spaces: 
\begin{equation}\label{eq.finitequotienttangent}
T_{[x]}\left(\bigslant{M}{G}\right)\cong\ \bigslant{T_xM}{T_x\left(G.x\right)}.
\end{equation}
This means, qualitatively, in $T_{[x]}(M/G)$ all directions showing ``orbitwards'' are modded out. To imitate this isomorphism "by construction" is the basic idea for the notion of tangent space at a point in $\Sh(M)$. The final definition, however, will then only be for $M=\R^n$.

The group of isometries $ISO(M)$ of a finite dimensional Riemannian manifold $M$ is a Lie group and in case $M=\R^n$, $ISO(M)=O(n)\ltimes\R^n$, where $O(n)$ is the orthogonal group of $\R^n$ and $\ltimes$ denotes the semidirect product of groups. 
We will find the above mentioned orbitward directions on Wasserstein space to be induced by the Lie algebra $\mathfrak{iso}(n)$ of $ISO(\R^n)$. One can show that $\mathfrak{iso}(n)=\mathfrak{so}(n)\ltimes \R^n$, where $\ltimes$ denotes the semidirect sum of algebras.

Recall from Section \ref{sec.dyn}, in particular Theorem \ref{them.ac}, that the tangent space on $W(M)$ is constructed using the weak continuity equation. A family of vector fields $v_t$ is considered to be tangent along an absolutely continuous curve $\mu_t$ if the weak continuity equation holds and if $\|v_t\|_{L^2(\mu_t)}$ is minimal among all the possible solutions. 
Now, to identify all those tangent vectors that show in the direction of the orbit, let us consider the pushforward of a measure $\mu$ by the flow of the left-invariant vector fields on $G=ISO(M)$. Without proof, we cite the following proposition. 

\begin{prop}\label{prop.fundconti}
	Let $X\in\mathfrak{g}$ and $\tilde{X}$ be the corresponding fundamental vector field on $M$. Further, let $\mu\in W(M)$ and $\mu_t:=\exp(-tX)_{\#}\mu$, $0\leq t\leq 1$. Then, the tuple $(\mu_t,\tilde{X})$ is a solution of the weak continuity equation \eqref{eq.wcont}, i.e. 
	\begin{equation}\nonumber
	\int_0^1\int_M\left(\frac{\partial}{\partial t}\varphi(x,t)+h(\nabla\varphi(x,t),\tilde{X}(x))\right)\ d(\exp(-tX)_{\#}\mu)(x)dt =0.
	\end{equation} 
\end{prop}

Let us note that because the tangent map of the orbit map $g_x:G\rightarrow G.x,\ g\mapsto g.x$ at point $e$, i.e. $(dg_x)_e:T_eG\rightarrow T_xG.x$, is surjective, every curve that is initially tangential to $G.x$ infinitesimally coincides with the integral line of a fundamental vector field starting at $x$. This is why in our case we will settle with those kinds of curves considered in Proposition \ref{prop.fundconti}.
By doing this, we naturally only consider the path-connected component of $Id_G$ with respect to which directions it is going to be modded out by.

The next question to be asked is whether $\mu_t$ from Proposition \ref{prop.fundconti} is indeed absolutely continuous in $W(M)$. According to Theorem \ref{them.ac} this is so, in case $\int_0^1 \|\tilde{X}\|_{L^2(\mu_t)}\ dt < \infty$. 

A positive answer for this is given in \cite{Lessel2025a} only in case $(M,h)$ is the Euclidean space. However, we repeat the conjecture that the same holds true also for compact manifolds and Riemannian manifolds with nonnegative curvature.

Unfortunately, $\tilde{X}$ is not necessarily the minimal vector field along $\mu_t$. But we can apply the orthogonal linear projection $P^\mu:L^2(\mu) \rightarrow  T_\mu W(\R^n)$, so that $P^{\mu_t}(\tilde{X})\in T_{\mu_t}W(\R^n)$ and the continuity equation still holds.
This is why we define
$$U_\mu:=\{P^{\mu}(\tilde{X})\mid \tilde{X} \text{ is a fundamental vector field} \}\subset L^2(\mu).$$ 
$U_\mu$ is a complete subspace of $T_\mu W(\R^n)$ and contains all the vectors pointing towards the orbit of $\mu$. With this definition, we propose the following notion for $T_{[\mu]}\Sh(\R^n)$.

\begin{dfn}[\emph{\textbf{Tangent space on $\Sh(\R^n)$}}]\label{dfn.tsh}
	We define the tangent space on $\Sh(\R^n)$ at point $[\mu]$ to be the set
	\begin{equation}\nonumber
	T_{[\mu]}\Sh(\R^n):=\ \bigslant{T_\mu W(\R^n)}{U_\mu}.
	\end{equation}
\end{dfn}

One can show \cite{Lessel2025a} that $\bigslant{T_\mu W(\R^n)}{U_\mu}\ \cong\ \bigslant{T_{g\mu} W(\R^n)}{U_{g\mu}}$. That is, definition \ref{dfn.tsh} is independent of the choice of the representative.

\begin{example}	
	\begin{equation}\nonumber
	T_{[\delta]}\Sh(\R^3)=\bigslant{\R^3}{U_\mu} \cong \{0\}.
	\end{equation}
	One may want to interpret this in the sense that there is no non-trivial classical one-particle motion in Shape space.
\end{example}

\subsection{Quantum dynamics on $\Sh(\R^n)$}\label{sec.final}
%---------------------------------------
% Kommentiere, warum es nicht schlimm ist, nur rotationssymmetrische Potetntiale zu betrachten. Ist das überhaupt so? Potentiale ändern die Symmetriegruppe der Mannigfaltigkeit! Man darf bei anwesenheit von Potentialen nur diejenige Untergruppe rausteilen, die auch das Potential invariant lässt. Lese dir dazu nochmal die Bemerkung von Mach auf Seite 51 durch!

This section now investigates the question in how far solutions of the free Schr\"odinger equation can naturally be regarded as curves in Shape space.\footnote{Let us point out that our aim here is different from the aim of investigating symmetry groups $\tilde{G}$ for the Madelung equations (or the Schr\"odinger equation) and from finding solutions that are invariant under the action of such a symmetry group and which therefore correspond to functions on the quotient space $M/\tilde{G}$. The reader who is interested in these kind of things should consult for example \cite{Olver2000}.}
 To this end, let $\psi(x,t)$ be a solution of the free Schr\"odinger equation (i.e. of equation \eqref{eq.schroe} with $V=0$), $\rho(x,t)=|\psi(x,t)|^2$ and $d\mu_t=\rho(x,t)d\lambda$, while $\lambda$ being the Lebesgue measure on $\R^3$.

\begin{prop}\label{prop.spreading}
	There is no $g\in ISO(\R^3)$ such that $\mu_t=g\mu_s$ for all possible times $s\neq t$. In particular, $\mu_t\neq\mu_s$ for all $t\neq s$.
\end{prop}
In particular, this means that $[\mu_t]\neq [\mu_s]$ for all times $t\neq s$, so that a passage from the curve of probability measures on the background space $\R^3$ to a curve of shapes (without explicit background) is possible in a meaningful and well-defined way.
\begin{proof}[Proof of Proposition \ref{prop.spreading}]
	To prove our claim we use formula (\ref{eq.spreading}), i.e. the fact that free wave packets spread over time. Let $t_0=0$ and $t>s$. Assume $\mu_t=g\mu_s$. Then according to Lemma \ref{lem.imagedensity}, $\rho(x,t)=\rho(g^{-1}(x),t)$.
	\begin{eqnarray}\nonumber
	\sigma^2_x(t) &=& \int_{\R^3}(x-\<x(t)\>)^2\ d\mu_t = \int_{\R^3}\left(x-\int_{\R^3}x\ d\mu_t\right)^2\ d\mu_t\\\nonumber
	            &=& \int_{\R^3}\left(x-\int_{\R^3}x\ \rho(g^{-1}(x),s)\ d\lambda(x)\right)^2 \rho(g^{-1}(x),s)\ d\lambda(x)\\\nonumber
	            &=& \int_{\R^3}\left(g(x)-\int_{\R^3}g(x)\ d\mu_s(x)\right)^2\ d\mu_s(x)\\\nonumber
	            &=& \int_{\R^3} \|g(x)\|^2\ d\mu_s(x) - \left(\int_{\R^3}g(x)\ d\mu_s(x)\right)^2
	\end{eqnarray}
	The last step is a general rule for variances. 
	
	It is $ISO(\R^3)=O(\R^3)\ltimes\R^3$, where $O(\R^3)$ is the orthogonal group of $\R^3$. So, suppose first that $g\in O(\R^3)$.
	Then, 
	\begin{eqnarray}\nonumber
	& & \int_{\R^3} \|g(x)\|^2\ d\mu_s(x) - \left(\int_{\R^3}g(x)\ d\mu_s(x)\right)^2\\\nonumber
	&=& \int_{\R^3} \|x\|^2\ d\mu_s(x) - \left(g\left(\int_{\R^3}x\ d\mu_s(x)\right)\right)^2\\\nonumber
	&=& \int_{\R^3} \|x\|^2\ d\mu_s(x) - \left(\int_{\R^3}x\ d\mu_s(x)\right)^2.
	\end{eqnarray}
	So $\sigma^2_x(t)=\sigma^2_x(s)$ which cannot be true due to the spreading of wave packets. 
	
	Assume now that $g(x)=x+a$, $a\in\R^3$. Then 
	\begin{eqnarray}\nonumber
	& & \int_{\R^3} \|g(x)\|^2\ d\mu_s(x) - \left(\int_{\R^3}g(x)\ d\mu_s(x)\right)^2\\\nonumber
	&=& \int_{\R^3} \left(\|x\|^2+2\<x,a\>+\|a\|^2\right)\ d\mu_s(x) - \left(\int_{\R^3}x\ d\mu_s(x)+\int_{\R^3}a\ d\mu_s(x)\right)^2\\\nonumber
	&=& \int_{\R^3}\|x\|^2\ d\mu_s(x)+\int_{\R^3}\<x,a\>\ d\mu_s(x)+\int_{\R^3}\|a\|^2\ d\mu_s(x)\\\nonumber
    & &	-\left(\int_{\R^3}x\ d\mu_s(x)\right)^2-\left(\int_{\R^3}a\ d\mu_s(x)\right)^2-2\<\int_{\R^3}x\ d\mu_s(x),\int_{\R^3}a\ d\mu_s(x)\>\\\nonumber
    &=& \sigma^2_x(s).
	\end{eqnarray}
	In this case, again, $\sigma^2_x(t)=\sigma^2_x(s)$ and thus a contradiction.
\end{proof}

\begin{corollary}
	Let $g\in ISO(\R^3)$. Then $\psi(x,t)\neq\psi(g^{-1}(x),s)$ for all $t,s$.
\end{corollary}

\begin{proof}
	Suppose $\psi(x,t)=\psi(g^{-1}(x),s)$, then also $|\psi(x,t)|^2=|\psi(g^{-1}(x),s)|^2$, i.e. $\rho(x,t)=\rho(g^{-1}(x),s)$, which does not hold true. 
\end{proof}

Let us now continue to study of the specific solution of the Schr\"odinger equation that we were already engaged with in section \ref{ch.quW}. In the following let $\psi(x,t)$ always be of the form \eqref{eq.Gaussfinal}, $\rho(x,t)$ of the form \eqref{eq.Gaussdensity}, $S(x,t)$ the phase of $\psi(x,t)$ and $F_s(\cdot,t)$ the flow map of $\nabla S(x,t)$ (see Lemma \ref{eq.flow}). Let furthermore $\mu_t$ be the measure defined by $d\mu_t=\rho(x,t)d\lambda(x)$.

\begin{lemma}\label{lem.o3}
	Let $g\in O(3)$. Then $\mu_t$ is invariant under the action of $g$, i.e. $g\mu_t=\mu_t$.
\end{lemma}

\begin{proof}
	\begin{eqnarray}\nonumber
	\rho(g^{-1}(x),t) &=& N^2(1+\frac{\hbar^2 t^2}{m^2l^4})^{-3/2} \exp\left(-\left(l^2+\frac{\hbar^2 t^2}{m^2 l^2}\right)^{-1}\|g^{-1}(x)\|^2\right)\\\nonumber
	&=& N^2(1+\frac{\hbar^2 t^2}{m^2l^4})^{-3/2} \exp\left(-\left(l^2+\frac{\hbar^2 t^2}{m^2 l^2}\right)^{-1}\|x\|^2\right)\\\nonumber
	&=& \rho(x,t).
	\end{eqnarray}
\end{proof}

In Theorem \ref{prop.opti} we have seen that $F_s(\cdot,t)$ is the optimal transport map from $\mu_s$ to $\mu_t$. Now we will see that the optimal transport of the ranged measure $g\mu_t$ to $\mu_s$ is given by the composition of moving $g\mu_t$ back to $\mu_t$ and transporting this to $\mu_s$.

\begin{prop}
	Let $g\in ISO(\R^3)$. Then $F_s(\cdot,t)\circ g^{-1}$ is the optimal transport map from $g\mu_s$ to $\mu_t$ for every $t,s$.
\end{prop}

\begin{proof}
	As in Theorem \ref{prop.opti}, we argue with the cost function $d^2/2$. (See also Remark \ref{rem.d^2/2}).	
		Because of Lemma \ref{lem.o3}, we only have to deal with isometries of the form $g(x)=x+a$, $a\in\R$. Let $g$ be such a map.
		Defining $C=\sqrt{\frac{1+\frac{\hbar^2}{m^2l^4}t^2}{1+\frac{\hbar^2}{m^2l^4}s^2}}$ and $\bar{\varphi}(x):=\frac{1}{2}C\|x-a\|^2$, 
		we see that $F_s(g^{-1}(x),t)=C(x-a)=\nabla \bar{\varphi}$. With this and $\varphi(x):=\frac{1}{2}\left(\|x\|^2-C\|x-a\|^2\right)$, the rest of the proof is analogous to the proof of Theorem \ref{prop.opti}.
\end{proof}

\begin{thm}
	Let $g(x)=x+a\in ISO(\R^3)$ with $a\in\R^n$. Then 
	\begin{equation}\label{eq.phythagoras}
	W(g\mu_s,\mu_t)^2= W(\mu_t,\mu_s)^2+\|a\|^2.
	\end{equation}
\end{thm}

Equation \eqref{eq.phythagoras} resembles Pythagoras' formula: Going with the flow of $\nabla S$ and translation by $a$ seem to be in this sense \emph{orthogonal} to each other.

\begin{proof} 
	Again, for calculational convenience we use the cost function $d^2/2$, so that in the end we have to multiply by the factor $2$.
	With the definition of $C$ as in the previous proof, we have
	\begin{eqnarray}\nonumber
	W^{d^2/2}(g\mu_s,\mu_t)^2 &=& \frac{1}{2}\int_{\R^3}\|x-F_s(g^{-1}(x),t)\|^2\ d(g\mu_s)(x)\\\nonumber
	                &=& \frac{1}{2}\int_{\R^3}\|g(x)-F_s(x,t)\|^2\ d\mu_s(x)\\\nonumber
	                &=& \frac{1}{2}\int_{\R^3}\|x-a-Cx\|^2\ d\mu_s(x).
	\end{eqnarray}
	Defining $\bar{Q}:=N^2\left(1+\frac{\hbar^2s^2}{m^2l^4}\right)^{-3/2}$ and $\bar{A}:=\left(l^2+\frac{\hbar^2s^2}{m^2l^2}\right)^{-1}$ transforms this into
	\begin{equation}\nonumber
	 \frac{\bar{Q}}{2}\int_{\R^3}\|(1-C)x-a\|^2\exp\left(-\bar{A}\|x\|^2\right)\ d\lambda(x).
	\end{equation}
	Performing a transformation with the diffeomorphism $\Phi(y)=\frac{1}{1-C}(y+a)$ yields
	\begin{eqnarray}\nonumber
	 & & \frac{\bar{Q}}{2}\int_{\R^3}\|y\|^2\exp\left(-\frac{\bar{A}}{(1-C)^2}\|y+a\|^2\right)\frac{1}{(1-C)^3}\ d\lambda(x)\\\nonumber
	 &=& \frac{\bar{Q}}{2(1-C)^3}\int_{\R^3}\|y\|^2\exp\left(-\frac{\bar{A}}{(1-C)^2}\left(\|y\|^2+\|a\|^2+2\<a,y\>\right)\right)\ d\lambda(x)\\\nonumber
	 &=& \frac{\bar{Q}}{2(1-C)^3}\exp\left(-\frac{\bar{A}}{(1-C)^2}\|a\|^2\right)\cdot\\\nonumber
	 & & \ \ \ \ \ \ \ \ \ \ \ \ \ \ \ \ \ \ \ \ \ \ \ \ \int_{\R^3}\|y\|^2\exp\left(-\frac{\bar{A}}{(1-C)^2}\left(\|y\|^2+2\<a,y\>\right)\right)\ d\lambda(x).
	\end{eqnarray}
	Define $A:=2\frac{\bar{A}}{(1-C)^2}$, $Q:=\frac{\bar{Q}}{(1-C)^3}\exp\left(-\frac{1}{2}A\|a\|^2\right)$ and $s:=-Aa$. We then have
	\begin{eqnarray}\nonumber
	& & \frac{1}{2}Q\int_{\R^3}\|y\|^2\exp\left(-\frac{1}{2}A\|y\|^2+\<s,y\>\right)\ d\lambda(x)\\\nonumber
	&=& \left(\frac{2\pi}{A}\right)^{3/2}\exp\left(\frac{1}{2A}\|s\|^2\right)\left(\frac{3Q}{2A}+\frac{Q}{2A^2}\|s\|^2\right)\\\nonumber
	&=& \left(\frac{2\pi}{A}\right)^{3/2}\exp\left(\frac{A}{2}\|a\|^2\right)\left(\frac{3Q}{2A}+\frac{Q}{2}\|a\|^2\right)\\\nonumber
	&=& \left(\frac{2\pi}{A}\right)^{3/2}\frac{Q}{2}\exp\left(\frac{A}{2}\|a\|^2\right)\left(\frac{3}{A}+\|a\|^2\right)\\\nonumber
	&=& \left(\frac{2\pi}{A}\right)^{3/2}\frac{\bar{Q}}{2(1-C)^3}\left(\frac{3}{A}+\|a\|^2\right)\\\nonumber
	&=& \left(\frac{2\pi(1-C)^2}{2\bar{A}}\right)^{3/2}\frac{\bar{Q}}{2(1-C)^3}\left(\frac{3(1-C)^3}{2\bar{A}}+\|a\|^2\right)\\\nonumber
	&=& \left(\frac{\pi}{\bar{A}}\right)^{3/2}\frac{\bar{Q}}{2}\left(\frac{3(1-C)^2}{2\bar{A}}+\|a\|^2\right).
	\end{eqnarray}
	Substituting the remaining abbreviations, we get
	\begin{eqnarray}\nonumber
	& & \pi^{3/2}l^3\left(1+\frac{\hbar^2 s^2}{m^2l^4}\right)^{3/2}\frac{(l^2\pi)^{-3/2}}{2}\left(1+\frac{\hbar^2 s^2}{m^2l^4}\right)^{-3/2}\cdot\\\nonumber
    & & \ \ \ \ \ \ \ \ \ \ \ \ \ \ \ \ \ \ \ \ \ \ \ \ \ \ \ \ \ \  \ \ \ \ 	    \left(\frac{3}{2}\left(l^2+\frac{\hbar^2s^2}{m^2l^2}\right)\left(1-\sqrt{\frac{1+\frac{\hbar^2t^2}{m^2l^4}}{1+\frac{\hbar^2s^2}{m^2l^4}}}\right)^2+\|a\|^2\right)\\\nonumber
	&=& \frac{1}{2}\left(\left(\frac{3}{2}l^2\left(1+\frac{\hbar^2s^2}{m^2l^4}\right)\left(1-\sqrt{\frac{1+\frac{\hbar^2t^2}{m^2l^4}}{1+\frac{\hbar^2s^2}{m^2l^4}}}\right)^2\right)+\|a\|^2\right)\\\nonumber
	&=& \frac{3}{4}l^2\left(\sqrt{1+\frac{\hbar^2s^2}{m^2l^4}}-\sqrt{1+\frac{\hbar^2t^2}{m^2l^4}}\right)^2+\frac{1}{2}\|a\|^2\\\nonumber
	&=& W^{d^2/2}(\mu_t,\mu_s)^2+\frac{1}{2}\|a\|^2.
	\end{eqnarray}
	The last step follows from Theorem \ref{prop.distance}.
\end{proof}

Evidently, $W(g\mu_s,\mu_t)$ is minimal, in case $a=0$. So we can formulate the following corollary.

\begin{corollary}
	For all times $t,s$ we have
	\begin{equation}\nonumber
	W_2(\mu_t,\mu_s)=D_2([\mu_t],[\mu_s]).
	\end{equation}$\hfill\square$
\end{corollary}

So not only is the passage from $\mu_t$ to $[\mu_t]$ a meaningful procedure, as we have seen in Proposition \ref{prop.spreading}, for this particular solution also the distances between the intermediate times do not change by this: There is no location on space a measure $\mu_t$ can be brought to such that cost for transporting it to any other $\mu_t$ gets cheaper. Or, in other words, $\mu_t$ cannot be brought closer to any $\mu_s$ than the location it is already situated.
Using Corollary \ref{cor.qugeod}, this leads to the following important statement.

\begin{thm}
	Let the curve of measures $\mu_t$ be defined by $d\mu_t=\rho(x,t)d\lambda(x)$, where $\rho(x,t)$ is of the form \eqref{eq.Gaussdensity}.
    Then for $s,t,u\in\R_{\geq0}$, $s<u<t$ it is 
    $$D([\mu_s],[\mu_t])=D([\mu_s],[\mu_u])+D([\mu_u],[\mu_t]).$$
    So $([\mu_t])$ is a geodesic in $\Sh(\R^3)$ in the sense of shortest paths.
\end{thm}
 
We have thus seen that the curve $\mu_t$ can naturally be regarded as motion in Shape space, as a curve that subjects the change of shapes. In particular without resorting to a spatial background structure. With the definitions in subsection \ref{s.shapespaces}, $\frac{1}{m}\nabla S(x,t)$ can furthermore naturally be regarded as tangent to $[\mu_t]$: Since $\mu_s\neq g\mu_t$ for any $t,s\in [0,\infty]$ and $g\in ISO(\R^3)$, $\nabla S$ can be attached to $[\mu_t]$ in a well-defined way. And because $\nabla S(x,t)\neq P^{\mu_t}(\tilde{X})$ for any fundamental vector field $\tilde{X}$, $[\nabla S(x,t)]\neq 0$.